\documentclass[runningheads]{llncs}
\usepackage[utf8]{inputenc}
\usepackage{amsmath}
\usepackage{amsfonts}
\usepackage{algorithm2e}
\usepackage{amssymb}
\usepackage[dvipsnames]{xcolor}
\usepackage{hyperref}
\usepackage{graphicx}
\usepackage{csquotes}
\usepackage[misc]{ifsym}
\usepackage[shadow]{todonotes}
\usepackage[nameinlink]{cleveref}
\Crefname{figure}{Fig.}{Figs.}
\usepackage{ek-todoboxes}
\usepackage{ek-goodies}
\DeclareGraphicsExtensions{.pdf}

\newcommand{\corrAuthor}{$^{\textrm{(\Letter)}}$}

\begin{document}

\title{Testing a Battery Management System via Criticality-Based Rare Event Simulation\thanks{The research leading to these results is funded by the German Federal Ministry of Education and Research (BMBF) within the project \enquote{TESTOMAT -- The Next Level of Test Automation} under grant agreement No. 01IS17026H.}
}

\subtitle{Extended Version\thanks{This is an extended version of the paper published by ACM that can be found at \url{https://doi.org/10.1145/3470481.3472701}.}}

\titlerunning{Testing a BMS via Criticality-Based Rare Event Simulation}
%
\author{Daniel Grujic\inst{1}\corrAuthor \and
Tabea Henning\inst{1} \and
Emilio José Calleja García \inst{2} \and \\
Andre Bergmann\inst{2}}
\authorrunning{D. Grujic, T. Henning, E.J. Calleja García, A. Bergmann}
%
\institute{%
	OFFIS e.V., Oldenburg, Germany\\
\email{\{thenning,dgrujic\}@offis.de}
\and AKKA Germany GmbH, Munich, Germany\\
\email{\{emilio-jose.calleja,andre.bergmann\}@akka.eu}
}
\maketitle              
\begin{abstract}%
For the validation of safety-critical systems regarding safety and comfort, e.g., in the context of automated driving, engineers often have to cope with large (parametric) test spaces for which it is infeasible to test through all possible parameter configurations. At the same time, critical behavior of a well-engineered system with respect to prescribed safety and comfort requirements tends to be extremely rare, speaking of probabilities of order $10^{-6}$ or less, but clearly has to be examined carefully for valid argumentation. Hence, common approaches such as boundary value analysis are insufficient while methods based on random sampling from the parameter space (simple Monte Carlo) lack the ability to detect these rare critical events efficiently, i.e., with appropriate simulation budget. For this reason, a more sophisticated simulation-based approach is proposed which employs optimistic optimization on an objective function called ``criticality'' in order to identify effectively the set of critical parameter configurations. Within the scope of the ITEA 3 TESTOMAT project (\url{http://www.testomatproject.eu/}) the collaboration partners OFFIS e.V. and AKKA Germany GmbH conducted a case study on applying \textit{criticality-based rare event simulation} to the charging process of an automotive battery management system given as a model. The present technical report documents the industrial use case, the approach, application and experimental results, as well as lessons learned from the case study.

\keywords{%
	Safety-Critical Systems \and
	Battery Management System \and
	Software Test Automation \and
	Rare Event Simulation \and
	Criticality \and
	Optimistic Optimization
}
\end{abstract}

\section{Introduction}\label{sec:intro}

Product development in the automotive industry is driven by the main vision of customer experience. This also applies to software development while safety standards and regulations such as ISO 26262~\cite{iso26262} and ISO/PAS 21448~\cite{isopas21448} have to be complied with as well. Both aspects are addressed during testing a system on various stages. Customer experience is usually confirmed by software tests and, at a latest step, by function designers and expert testers. However, the subjective evaluation of the functionality tends to be highly error-prone and incomplete, and thus has to be augmented by automated test methods. In this context, we propose and apply exemplarily a novel simulation-based approach to analyze and evaluate software systems that may comprise (rare) safety-critical events. 

For the safety validation of those safety-relevant applications, engineers often have to cope with large parametric test spaces, i.e., high-dimensional continuous parameter spaces, for which it is infeasible to test through all possible parameter configurations. The approaches to overcome this difficulty are manifold, comprising methods such as boundary value analysis and equivalence testing. However, these approaches have in common that they assume the test space to be structured in a certain way regarding the (critical) behavior of the considered system. For this reason, the common sampling-based approach of statistical model checking~\cite{Legay2010} can be applied to infer statistically whether the system gives evidence to satisfy some prescribed safety requirements that are potentially of probabilistic nature. To this end, one could consider using simple Monte Carlo simulation~\cite{Morio2014}, meaning that samples from the given parameter space are generated randomly (uniformly) according to the original probability distribution defined on the space. In the context of testing well-engineered safety-critical systems, however, the event that the safety requirements are violated is typically extremely rare (speaking of probabilities of order $10^{-6}$ or less)~\cite[p. 14]{Puch2019diss}. Thus, applying simple Monte Carlo leads to excessively high simulation effort required to observe these rare events during simulation. For instance, it can be shown that for an event with a probability of order $10^{-4}$ at least $10^{6}$ samples are needed to estimate this probability with a relative error of $10\%$~\cite{Morio2014}.

In order to reduce the simulation effort needed to detect such rare critical events, more sophisticated guiding approaches can be used that require an objective function. We refer to this as \textit{criticality function} for which high values indicate the occurrence of critical situations w.r.t. the given requirements. The idea is then to apply the guiding approaches to optimize this function in order to find critical parameter configurations, i.e. those that lead to harmed requirements.

\vspace*{-0.7em}\paragraph{Contribution}
Within the scope of this case study, we applied and evaluated algorithms from the domain of \textit{optimistic optimization} \cite{bartlett18}, \cite{munos} to the charging process of an automotive battery management system given as a model provided by AKKA in the context of the ITEA 3 TESTOMAT project\footnote{\url{http://www.testomatproject.eu/}}. The overall goal of the case study is to showcase the efficacy of a simulation-based methodology developed by OFFIS in the sense that (rare) critical events in the system are detected with a notably increased frequency compared to simple Monte Carlo simulation. In this way, it is possible to efficiently identify critical parameter configurations, i.e., test cases, that can be used, e.g., for the purpose of test prioritization within a test suite. 

\vspace*{-0.7em}\paragraph{Outline} 
This technical report is structured as follows. In Section~\ref{sec:testsystem}, we start with a description of the test system. In Section~\ref{sec:methodology}, the overall methodology is explained. Section~\ref{sec:application} includes the application and experiments that have been conducted, and Sections~\ref{sec:discussion} and~\ref{sec:conclusion} conclude the results.

\section{Test System}\label{sec:testsystem}
\begin{figure*}[h!]
	\centering
	\includegraphics[width=0.75\textwidth]{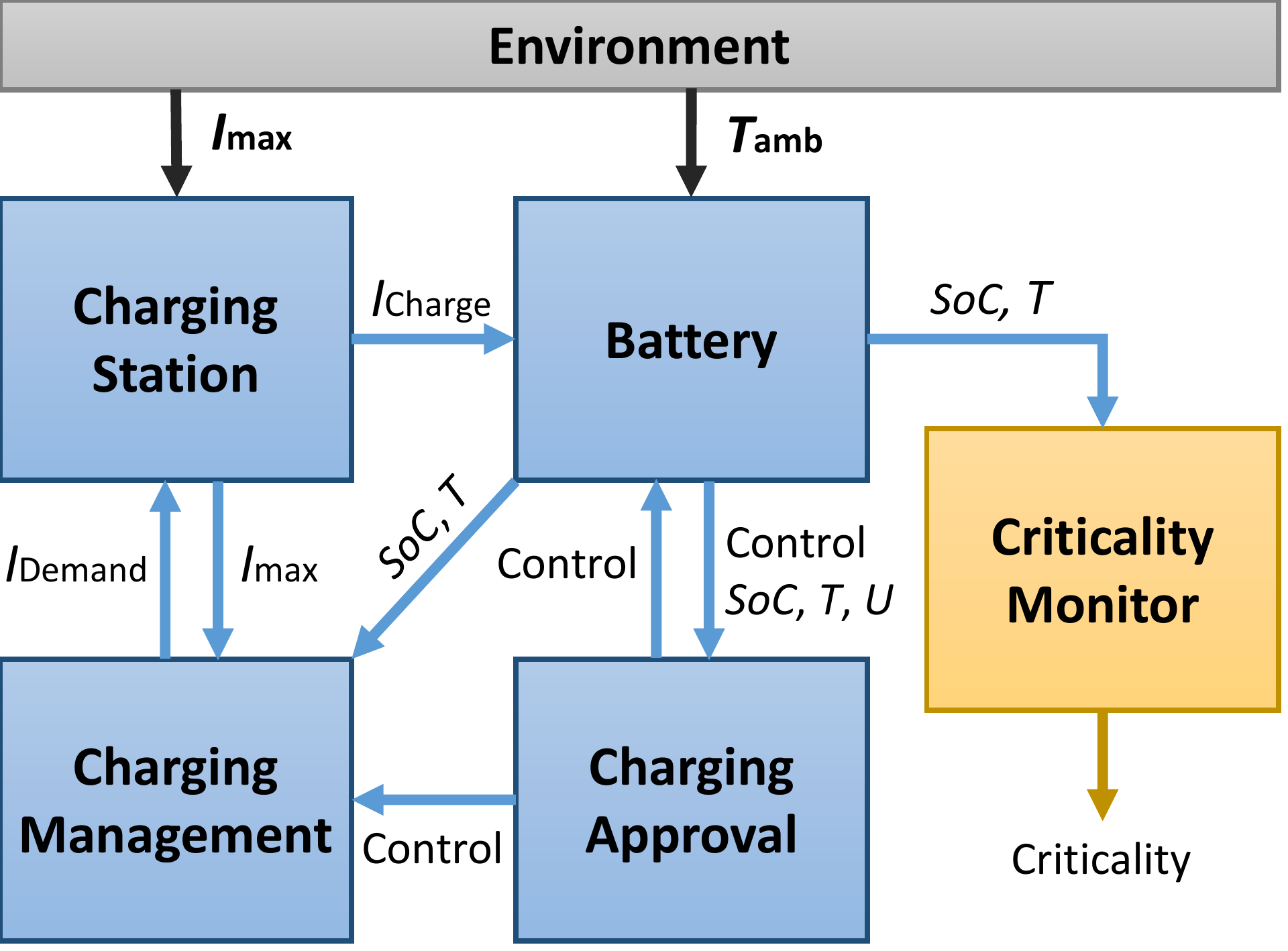}
	\caption{
		Overview of the overall system. The \textcolor{RoyalBlue}{blue} rectangles represent the components of the system under test. The \textcolor{Gray}{gray} box represents the environment and the \textcolor{Dandelion}{yellow} box is a monitor that observes the current state of the battery. The arrows depict the flow of signals between the different components. The full names of the signals can be found in Tab.~\ref{tab:signals}. The \textbf{bold} signals are input parameters to the system and constant over the course of a single simulation.
	}\label{fig:test-system}
\end{figure*}

Our test object is the model of a charging process that controls the state of charge of an industrial battery model. The overall \textit{system under test} (SUT) consists of four components: the battery to be charged, the charging station, a charging approval component starting and stopping the charging process, and a charging management component that decides which current will be delivered by the charging station. An overview of the system can be found in Figure~\ref{fig:test-system}. The full system depicted there is denoted as the use case in the following sections.

\begin{table}
	\begin{center}
		\begin{tabular}{ l | l }
			\; \textbf{Name [Unit]} \; & \; \textbf{Full Name} \\
			\hline 
			\; $\boldsymbol{I_{\textbf{max}}}$ [A] \; & \; Maximum Available Current  \\
			\; $I_{\text{charge}}$ [A] \; & \; Charging Current  \\
			\; $I_{\text{demand}}$ [A] \; & \; Demanded Current  \\
			\; $\boldsymbol{T_{\textbf{amb}}}$ [°C] \; & \; Ambient Temperature  \\
			\; $T_{\text{bat}}$ [C°] \; & \; Battery Temperature  \\
			\; $U_{\text{bat}}$ [V] \; & \; Battery Voltage  \\
			\; $SoC$ \; & \; Battery State of Charge  \\
			\; \textit{Control} \; & \; Control Signal  \\
		\end{tabular}
	\end{center}
	\vspace{0.3cm}
	\caption{Overview of the signals in the system. The \textbf{bold} signals are input parameters to the system and constant over the course of a single simulation.}\label{tab:signals}
\end{table}

In general the system works as follows: the battery is charged by the charging station. This charging process is monitored by the charging approval and the charging management functions. The charging management function specifies the current that is used to charge the battery, and the charging approval function starts and stops the charging process depending on the state of charge of the battery, $SoC$, and its temperature $T_{\text{bat}}$ by sending a specific control signal. There are two parameters that can be considered as open input from the environment and thus directly influence the system: (i) the ambient temperature $T_{\text{amb}}$ cooling the battery, and (ii) the maximum available current $I_{\text{max}}$ which corresponds to the maximum current that can be provided for the charging process by an electrical grid. In this model both parameters are assumed to remain constant during the charging process.

The component on the right in Figure \ref{fig:test-system} is the criticality monitor assessing the system under test with regard to the specified requirements that can be found below. For detailed information regarding the criticality we refer the reader to Sections~\ref{sec:methodology} and~\ref{sec:application}.  

A more detailed description of the different components is given in the following subsections.

\subsection{Battery}\label{subsec:battery}
\begin{figure*}[h!]
	\centering
	\includegraphics[width=\textwidth]{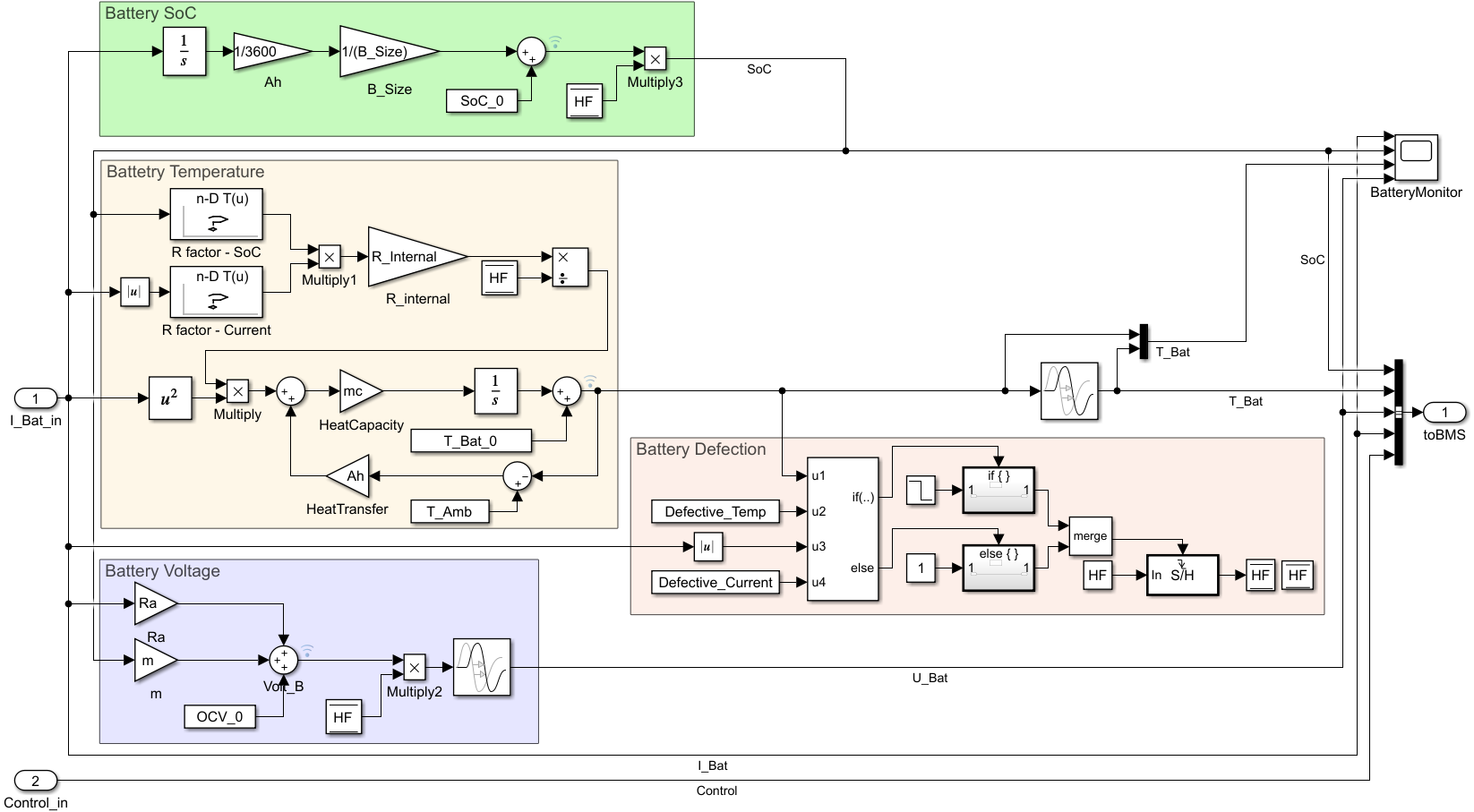}
	\caption{Simulink battery model.}\label{fig:battery-model-simulink}
\end{figure*}

The battery model consists of four different modules: battery SoC, battery temperature, battery voltage and battery defection, which are depicted in Figure \ref{fig:battery-model-simulink}.\\

\hspace*{-0.55cm}\underline{Battery SoC}
\vspace*{0.1cm}

\hspace*{-0.55cm}In the \textit{battery SoC} module the coulomb counting method is implemented for the calculation of the $SoC$. It is also known as ampere hour counting and current integration, and is a standard technique for calculating the $SoC$ (see e.g. \cite{murnane2017}, \cite{piller2001}). 
In order to calculate the \textit{SoC} values, readings of the battery current are integrated mathematically over the usage period $\tau$:

$$ SoC(t_0 + \tau) = SoC(t_0) + \frac{1}{B_\text{size}}\int_{t_0}^{t_0+\tau} I_\text{charge}(t) \,dt \,,$$
where
\begin{itemize}
	\item[--] $SoC(t_0) = SoC_0$ is the initial $SoC$ at the beginning of the charging process,
	\item[--] $B_\text{size}$ is the battery capacity (in $\text{Ah}$),
	\item[--] $I_\text{charge}(t)$ is the charging current of the battery and corresponds to $I_{bat}$ in the module. It is delivered by the charging station (see Fig.~\ref{fig:test-system}).
\end{itemize}
 The method (\cite{murnane2017}, \cite{piller2001}) calculates the remaining capacity by accumulating the charge transferred in or out of the battery (here: in). 
The coulomb counting method is straightforward, but its accuracy strongly depends on a precise measurement of the battery current as well as on an accurate estimation of $SoC(t_0)$. Hence, if a pre-known capacity is at hand (e.g., memorized or initially estimated by the operating conditions), the method can be applied without enhancements to calculate \textit{SoC} values. However, losses during charging (and discharging) and effects of self-discharging will cause accumulating errors, leading to potentially imprecise \textit{SoC} estimation. We neglect these factors in our case for the sake of simplicity.\\  

\hspace*{-0.55cm}\underline{Battery Temperature}
\vspace*{0.1cm}

\hspace*{-0.55cm} The \textit{battery temperature} module is implemented according to the following differential equation for $T_{bat}$ which is a simplified version of the governing equation presented in \cite{Ismail_2013}.

$$ m \cdot c_\text{cell} \frac{dT_\text{bat}}{dt} = R \cdot I_\text{charge}^2 + A \cdot h \cdot (T_\text{amb} - T_\text{bat}) \,,$$
where 
\begin{itemize}
	\item[--] $m$ is the total mass of the battery (in kg) and $c_\text{cell}$ is the specific heat capacity of the battery (in J/(kg·K)). The heat capacity $mc$ is the inverse product of both,
	\item[--] $R$ is the internal resistance of the battery in Ω. It is the product of the variable \textit{R\_internal} and some \textit{R factors} that depend on the \textit{SoC} and the current $I_{charge}$ (see Fig. \ref{fig:battery-model-simulink}).
	\item[--] $A$ is the surface area of the battery (in $\text{m}^2$) on which it can lose or absorb heat from the environment and $h$ is the heat transfer coefficient (in W/($\text{m}^2$·K)), which depends on the cooling system. The heat transfer $Ah$ is the multiplication of both.
	\item[--] $T_\text{amb}$ denotes the ambient temperature introduced above
\end{itemize}

\hspace*{-0.55cm}\underline{Battery Voltage}
\vspace*{0.1cm}

\hspace*{-0.55cm}Within the \textit{battery voltage} module, a common linear approximation model \cite{chang2013} is used for the calculation of the battery voltage $U_\text{bat}$ which is the result of the following equation:
$$U_\text{bat} = R_a \cdot I_\text{charge} + m \cdot SoC + OCV_0\,,$$

where $R_a$ is the pre-resistance of the battery in Ω. The constants $m$ (slope) and $OCV_0$ (open circuit voltage at 0\% SoC) are derived from experimental measurements where a linear estimation of the open circuit voltage $OCV$ against $SoC$ is performed.\\

\hspace*{-0.55cm}\underline{Battery Defection}
\vspace*{0.1cm}

\hspace*{-0.55cm}In addition, the present battery model includes a \textit{battery defection} module that simulates damaged battery cells. In the context of this case study we restrict ourselves to using a healthy battery to exclude further sources of error which means that the \textit{battery defection} module is not used.

\subsection{Charging Approval}\label{subsec:charging-approval}
\begin{figure*}[h!]
	\centering
	\includegraphics[width=\textwidth]{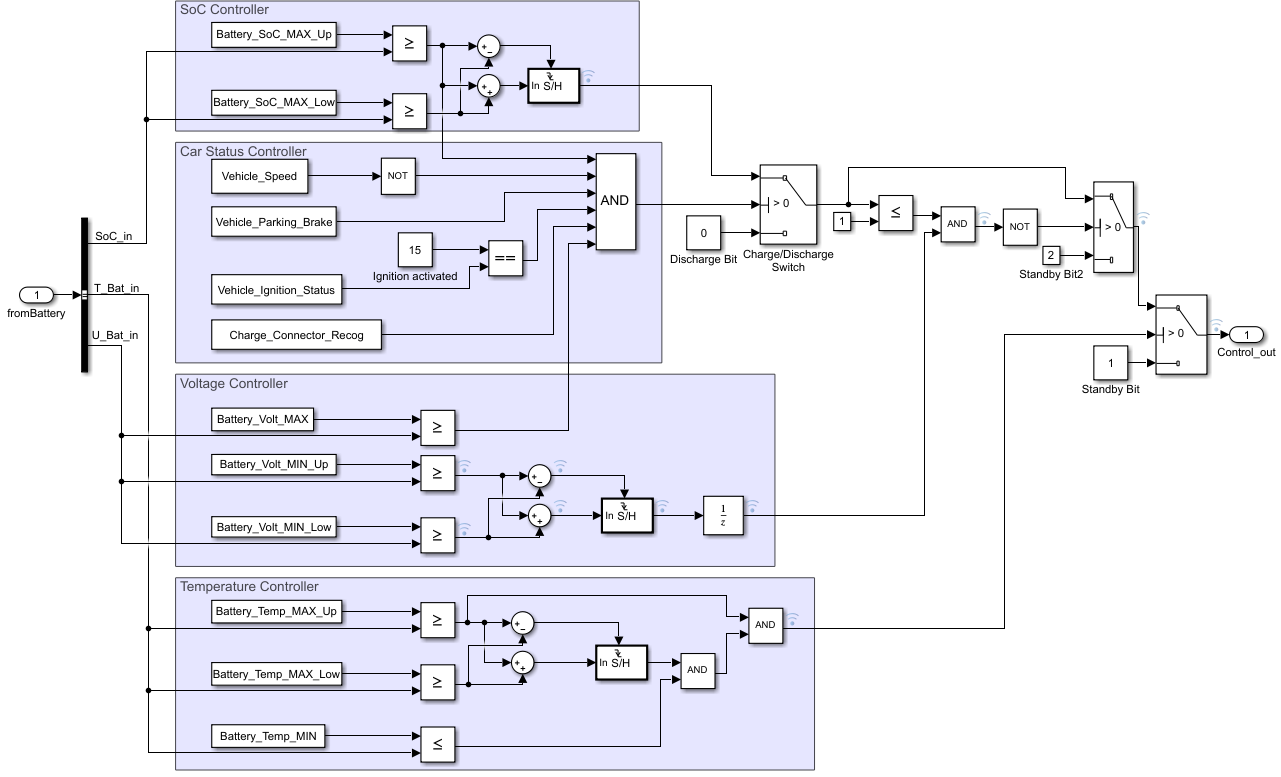}
	\caption{Charging approval model in Simulink.}\label{fig:charg-appr-simulink}
\end{figure*}

The approval of charge depends on the following parameters: the battery $SoC$, the battery voltage $U_{bat}$, the battery temperature $T_{bat}$, and the car status as shown in Figure~\ref{fig:charg-appr-simulink}. 

The charging approval output, respectively called \textit{Control} (see Fig.~\ref{fig:test-system}), decides about the current charging state which can be ``charging'', “discharging” or “resting”. The charging process will be enabled if the following three conditions are fulfilled:
\begin{itemize}
	\item[--] The car status is ready for charge which means that the vehicle is stopped, the parking brake activated, the charge connector is recognized and ignition is activated. (In the context of our case study these conditions are always fulfilled.)
	\item[--] The battery is not fully charged, meaning that the $SoC$ of the battery is lower than $95\%$.
	\item[--] The temperature $T_\text{bat}$ and the voltage of the battery do not exceed certain minimum and maximum values. 
\end{itemize}
When the battery is fully charged ($SoC \geq 95\%$), \textit{Control} is set to discharging, i.e., the charging process is accomplished and the battery is ready to be discharged.

When the third condition is not met, the battery will switch to a resting mode for which the charging process is interrupted until the above conditions are fulfilled again. This condition is intended to ensure that the battery will neither overheat nor overcharge. One drawback of these interruptions is that the charging time will increase, which is an effect that has to be tested. 

Additionally, for a more realistic behavior of the module, a control of hysteresis has been implemented for the battery $SoC$, battery voltage and battery temperature.

\subsection{Charging Management}\label{subsec:charging-management}
\begin{figure*}[h!]
	\centering
	\includegraphics[width=\textwidth]{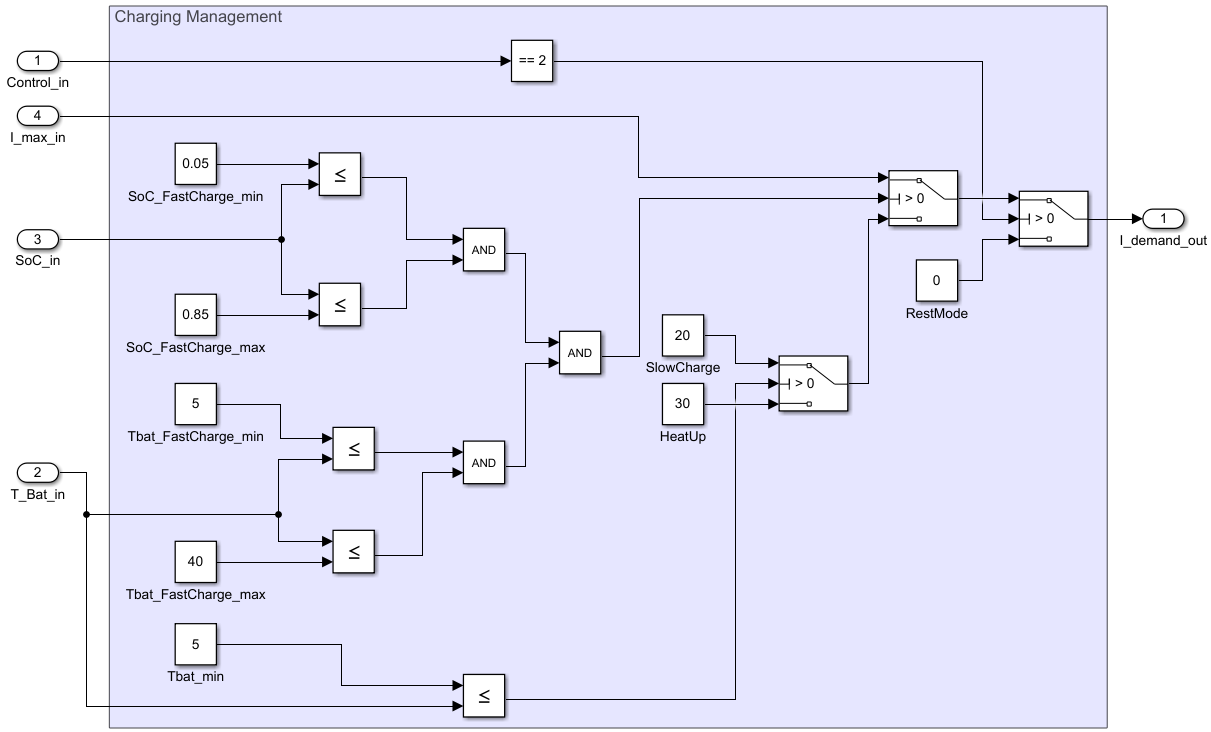}
	\caption{Charging management model in Simulink.}\label{fig:charg-manage-simulink}
\end{figure*}

The Charging Management System (see Figure \ref{fig:charg-manage-simulink}) decides which current is delivered to the battery. Depending on the state of charge $SoC$ and the temperature of the battery $T_\text{bat}$, it will choose between four different charging modes characterized by a specific requested current $I_\text{demand}$:
\begin{enumerate}
	\item \textbf{Heat Up} [$I_\text{demand}=30$ A]: This charge mode is used when the temperature of the battery is very low in order for the battery to heat up. It is used whenever the temperature of the battery falls below $5$ °C.
	\item \textbf{Fast Charge} [$I_\text{demand}=I_\text{max}$]: When using fast charge, the battery will be charged with the maximum available current to keep the charging time as small as possible. This mode is used when the state of charge is between 5\% and 85\% and the temperature lies within the range of normal operation for the battery which is between 5 °C and 40 °C.
	\item \textbf{Slow Charge} [$I_\text{demand}=20$ A]: The slow charge mode has the purpose of protecting the battery under stress conditions. This mode is used whenever the conditions for heat up and fast charge are not fulfilled. 
	\item \textbf{Rest Mode} [$I_\text{demand}=0$ A]: This mode is used when the Charging Approval System interrupts the charging process. No current is delivered in this mode.  
\end{enumerate}

\subsection{Charging Station}\label{subsec:charging-station}
The Charging Station delivers the charging current $I_\text{charge}$ to the battery. The charging current corresponds to the demanded current unless the demanded current $I_\text{demand}$ is greater than the maximum available current $I_\text{max}$ such that $I_\text{charge}=\min\lbrace I_\text{demand}, I_\text{max}\rbrace$.

\subsection{Requirements and Test Space}\label{subsec:reqs-and-test-space}
The charging process naturally causes the battery to heat up. As this can lead to faster aging and/or defects in the battery cells, high temperatures have to be avoided. At the same time, the battery should also be charged as fast as possible which demands a high current leading to a stronger heating of the battery. The following requirements are given:
\begin{enumerate}\itshape
	\item The charging time must not exceed 7.2 h! 
	\item The temperature of the battery must not exceed 51 °C!
\end{enumerate}
These requirements should hold for ambient temperatures between $-5$ °C and $40$ °C, and for charging currents between 10 A and 100 A, for a given battery type with fixed properties. This means that, in our case, the \textit{test space} is given by the Cartesian product of the following intervals (value ranges of real numbers):
\begin{enumerate}
	\item $T_\text{amb}$ [in °C]: $[-5, 40]$,
	\item $I_\text{max}$ [in A]: $[10, 100]$.
\end{enumerate}

\section{Methodology}\label{sec:methodology}
\textit{Criticality-based rare event simulation} represents an approach that OFFIS continuously develops, particularly for the simulation-based safety validation of safety-critical systems. This approach is mainly inspired by rare event simulation~\cite{Morio2014} which aims at increasing the chance to observe rare (critical) events during the simulation of the SUT in its environment. The idea behind the approach is to use a guiding function based on expert or domain knowledge as input instead of randomly drawing values from a given parameter set.
Furthermore, by incorporating learning methods, the results of completed simulation runs can be taken into account when selecting future simulation parameters and thus guide the system into critical situations faster. In the context of automated driving, this principle idea has been applied for the purpose of safety validation, e.g. in~\cite{okelly2018},~\cite{zhao16} and~\cite{zhao17}.

The methodology requires a real-valued parameter space $X$ (the test space) ---reflecting the environment of the SUT--- to be given, where each parameter point $x\in X$ represents a combination of single input configurations for the simulation of the system. Further, a user of our methodology has to be able to determine a numerical, finite criticality value for every realized parameter sample $x \in X$ with regard to the imposed (safety) requirements, i.e., we assume a user-defined numerical, bounded, deterministic \textit{criticality function} $\kappa$ is given, which is defined on the parameter space $X$ and reflects the degree of violated requirements. Moreover, a suitable criticality threshold $c_\kappa$ needs to be determined so that the \textit{critical event} $\lbrace\kappa \ge c_\kappa\rbrace$ coincides with the requirements being violated. 

Now, the goal of the proposed framework is to “cover” the set of critical parameter instances $C_\kappa$ as well as possible by running simulations according to a prescribed simulation budget $N$ (number of simulations). In other words, we aim at identifying as much critical events as possible among the simulation runs. As mentioned earlier, it appears inadequate to simply sample random parameter points from the parameter space if the critical event is assumed to be very rare, speaking of probabilities of order $10^{-6}$ or even less~\cite{Morio2014}. This would correspond to applying so-called \textit{simple Monte Carlo simulation}.  

Within the scope of the present case study, the main goal is to apply several algorithms from the domain of \textit{optimistic optimization} (OO) \cite{bartlett18}, \cite{munos} to the system under test and compare the results with those obtained from running simple Monte Carlo simulations in order to provide evidence that OO is more adequate for detecting and covering rare critical events when given the same simulation budget. For this purpose, we make use of the fact that, in contrast to “classical” optimization, OO approaches aim to optimize a given unknown, real-valued objective function, i.e., the criticality function in our context, globally over some feasible region, while requiring minimal assumptions regarding the properties of the function~\cite{bartlett18}. Within our methodology, however, we are not interested in “only” finding a global maximum of the criticality. Instead, we employ \textit{OO-guided simulation} to observe rare critical events efficiently as follows. First, a specific OO algorithm (see below) is chosen and instantiated by determining the corresponding hyperparameters (if any), along with the simulation budget $N$. In order to obtain criticality values $\kappa(x_i), i=1,...,N$, the following iteration loop is carried out. According to the chosen OO algorithm, single parameter configurations $x_i \in X$ are generated iteratively and passed as input to a simulation engine, e.g., simulating the combined system [SUT+environment] for the specified parameter configuration. Subsequently, the corresponding criticality values $\kappa(x_i)$ are each observed by a criticality monitor for the simulation run induced by the parameter sample $x_i$. The criticality monitor might also observe whether the threshold $c_\kappa$ is exceeded, i.e., whether a critical event occurs. In a next step, the observed criticality value is passed back to the OO test manager that derives the succeeding parameter configuration $x_{i+1}$ according to the applied OO algorithm. This realization then serves as input for the simulation, and so forth, until the simulation budget $N$ is exhausted. 

Loosely speaking, the OO algorithms that we focus on in this case study iteratively divide the search space $X$ into cells, providing a partition of $X$ whose resolution is assumed to be finer in regions where the criticality function takes large values. Thus, compared with simple Monte Carlo simulation, we expect to obtain a higher degree of coverage of the critical space $C_\kappa$, i.e., to detect considerably more critical events $\lbrace\kappa \ge c_\kappa\rbrace$ when applying any OO algorithm instead. On a technical level, an OO algorithm incrementally builds a hierarchically structured $K$-ary tree\footnote{Within the scope of the present case study, we throughout use binary trees.} of subregions $\{P_{h,i}\}$ that represents a disjoint partition of the parameter space at every depth $h\ge0$, where each node $(h,i)$ ($h$: depth, $i$: index) is associated with a specific subregion of $X$. Whenever a node is split, the corresponding region $P_{h,i}$ is divided into $K$ subsets. This is represented by creating $K$ children nodes from node $(h,i)$ which are located at the next depth level in the tree, i.e., at level $h+1$. Based on the assumptions, some OO approaches compute optimistic upper bounds (often referred to as the $B$-values of the nodes) on the maximal function value for each of the subregions to indicate how profitable it is to choose the corresponding node. 

In the course of the present case study, we consider the following approaches (the OO algorithms are briefly outlined below): 
\begin{enumerate}
	\item Simple Monte Carlo Simulation 
	\item Hierarchical Optimistic Optimization (\texttt{HOO})
	\item Parallel Optimistic Optimization (\texttt{POO})
	\item Deterministic Optimistic Optimization (\texttt{DOO})
	\item Simultaneous Optimistic Optimization (\texttt{SOO}).
\end{enumerate}

\subsection*{Hierarchical Optimistic Optimization (HOO)}
The \texttt{HOO} algorithm introduced in~\cite{Bubeck2011} is a popular multi-armed bandit strategy and is initially conceived for noisy optimization, meaning that function evaluations can be distorted by bounded noise. In order to apply \texttt{HOO}, the criticality $\kappa$ is assumed to be weakly Lipschitz (continuous) w.r.t. a so-called dissimilarity function that is assumed to locally set a bound on the decrease of the function around each of its global maxima. This involves the existence of two smoothness (hyper-)parameters $\nu_1>0$ and $0<\rho<1$ that have to be chosen suitably so that a set of assumptions is guaranteed to hold (cf. \cite[Section 4.1]{Bubeck2011}). More precisely, these hyperparameters are part of the so-called $U$-value, an initial estimate of the maximum of the mean-payoff function (i.e. the criticality) in a region $P_{h,i}$ associated with a node $(h,i)$:\footnote{This definition will be useful with a view to performance evaluation of \texttt{HOO}.}
\begin{equation}\label{eq:U-value}
U_{h,i}(N) = \hat{\mu}_{h,i}(N) + \sqrt{\frac{2 \ln N}{T_{h,i}(N)}} + \nu_1 \rho^h\,,
\end{equation}
if $T_{h,i}(N)$, the number of times a descendant of $(h,i)$ is selected up to round $N$, is greater than zero (otherwise, $U_{h,i}(N) = +\infty$). Here, the first and second term account for the average of rewards and the uncertainty arising from the randomness of the rewards, respectively. The third term, $\nu_1 \rho^h$, accounts for the maximum possible variation of the objective function over the region $P_{h,i}$ and decreases exponentially with increasing depth of the tree $h$. We note that the $U$'s are auxiliary values for the subsequent computation of the $B$-values, which in turn drive the decisions along the path in the tree to select the next node to “play” (evaluate a point in the corresponding region). Thus, as $\nu_1$ and $\rho$ are the only free parameters in the calculation of $U$-values, they have to be chosen carefully in order for \texttt{HOO} to perform well. 

As mentioned above, in every round of the algorithm the current tree of partitions (i.e., the nodes) is searched by selecting the most promising child at every depth according to the $B$-values until a leaf node is reached. In the associated subregion, a parameter point $x_i$ is drawn randomly and simulated. The corresponding “reward” $\kappa(x_i)$ is evaluated and the statistics stored in the tree of selected nodes as well as the $B$-values are updated (backwards).

\subsection*{Parallel Optimistic Optimization (POO)}
It can be shown that the performance of \texttt{HOO} heavily depends on the knowledge of the smoothness parameters, which in turn requires a good knowledge of the behavior of the criticality around its global optima. For this reason, the \texttt{POO} algorithm has been introduced in~\cite{Grill2015}. The core idea is to run several \texttt{HOO} instances with varying choices of $\nu_1$ and $\rho$ in parallel, to propagate the obtained function evaluations among the instances and to select the best performing instance. 

The \texttt{POO} algorithm requires as input the smoothness parameters $\nu_{\text{max}}>0$ and $0<\rho_{\text{max}}<1$ being respectively a factor and an approximately upper bound on the corresponding inputs for the \texttt{HOO} instances to be tested. More precisely, \texttt{POO} starts \texttt{HOO} instances using $\nu_{\text{max}}$ and varying exponents of $\rho_{\text{max}}$ as inputs. The notion of $\nu_{\text{max}}$ and $\rho_{\text{max}}$ is to control the trade-off between performance (optimizing the function with given smoothness parameters) and size of the comparison class, i.e., the larger the values of the hyperparameters used by \texttt{POO}, the larger the set of \texttt{HOO} instances to run and compare (cf.~\cite[Appendix C]{Grill2015}).

\subsection*{Deterministic Optimistic Optimization (DOO)}
Both \texttt{HOO} and \texttt{POO} are suited for optimizing functions with possibly noisy function evaluations, e.g., due to sensor inaccuracy. Hence, assuming the evaluation of the criticality function to be noiseless and deterministic, we can expect that noisy optimization algorithms will perform sub-optimally when applying them for our purposes. We thus take deterministic OO algorithms into consideration within the scope of the present case study.

One of those approaches is the \texttt{DOO} algorithm as introduced in~\cite{munos}. In \texttt{DOO}, each node $(h,i)$ of the tree gradually built by the algorithm is associated with a specific (fixed) point $x_{h,i} \in X$, e.g., the center point of the associated subregion $P_{h,i}$. In every round $t$, \texttt{DOO} selects from all current leaves that one with the highest $b$-value defined as 
$$b_{h,i} = \kappa(x_{h,i}) + \delta(h),$$
where the sequence $\delta(h)$ is the only input of the algorithm. Having determined that leaf, the algorithm splits it, i.e., $K$ child nodes are created, for each of which the specific point in the associated region is evaluated and its $b$-value is computed in a last step of the round. 

Similarly to the before mentioned approaches, \texttt{DOO} requires the criticality function to be locally smooth w.r.t. a so-called semi-metric $l$ around at least one of its global optima $x^*$ (cf.~\cite[Assumption 2]{munos}). Intuitively, this assumption means that the criticality may not decrease too fast around one of its global maxima, which represents some sort of a locally one-sided Lipschitz condition. Besides that, the hierarchical partitioning has to admit bounded diameters and well-shaped subregions w.r.t. the decreasing sequence $\delta(h)$ in terms of $l$ (cf.~\cite[Assumptions 3-4]{munos}). In particular, the semi-metric $l$ has to be known to the user in order to apply the \texttt{DOO} algorithm effectively.

\subsection*{Simultaneous Optimistic Optimization (SOO)}
In case the semi-metric $l$ is unknown in the above context, the \texttt{SOO} algorithm~\cite{munos} can be used while assuming the same assumptions as for \texttt{DOO} to hold. The core idea for \texttt{SOO} is to select the leaf with the highest associated reward $\kappa(x_{h,i})$ at each depth $h$ in every round, and to split it if its reward exceeds all previously observed rewards in that round. This process is kept until either the bottom of the current tree or a specific threshold $h_\text{max}(t)$ is reached, which is a user-given input pruning the tree at maximal depth after $t$ node splittings. 
The input function $h_\text{max}(t)$ is intended to control the behavior of the algorithm regarding the trade-off between exploration and exploitation, and thus has to be chosen carefully. Large values for $h_\text{max}$ can lead to deep trees such that regions with high rewards in the past are sampled more often, while small values force the algorithm to sample preferably in less explored regions.

\section{Application and Experiments}\label{sec:application}
\subsection{Modeling the Criticality Function}\label{subsec:modeling-crit-function}
In order to apply the above-outlined methodology on the present use case, we have to model the given requirements into an adequate criticality function $\kappa$ that is defined on the space of admissible parameter configurations (for requirements and parameter space, see Sec.~\ref{subsec:reqs-and-test-space}). For this, it has to be taken into account that the criticality function is required to be a numerical, bounded function (without loss of generality taking values between 0 and 1) which is well-defined on the whole parameter space, i.e., assigning a unique value for every specified parameter point $x \in X = [-5~^\circ\text{C}, 40~^\circ\text{C}] \times [10~\text{A}, 100~\text{A}]$ (see Sec.~\ref{subsec:reqs-and-test-space}). Additionally, the criticality function has to correlate loosely with the degree of violated requirements. That is, the higher the criticality value for a given parameter point, the more the requirements are harmed, and vice versa. In particular, it should be true that the criticality function hits the criticality threshold $c_\kappa$ if and only if at least one of the given requirements regarding the charging time and the battery temperature is violated. For the present use case, a criticality threshold of $c_\kappa=0.8$ is prescribed such that the range of critical values (exceeding the threshold) is adequately broad to differentiate between “only just critical” parameters hitting the threshold and those for which the maximum criticality $\kappa=1$ is reached. The latter case corresponds to exceeding the requirement thresholds in a way that at least one of the properties (charging time/battery temperature) takes some unacceptable, fatal value that needs to be avoided in any case. That is, we have to scale the given values from the requirements ($t_\text{charge}=7.2~\text{h}$  and $T_\text{bat}=51~^\circ\text{C}$, respectively), which correspond to $c_\kappa=0.8$, to \textit{fatal values} that will in turn correspond to reaching the maximum criticality. Scaling yields the fatal values
\begin{equation}\label{eq:fatal-values}
t_\text{charge,fatal} = 9~\text{h} \quad \text{and} \quad  T_\text{bat,fatal} =  63.75~^\circ\text{C}\,.
\end{equation}
The basic idea for the criticality function used is to define an auxiliary function for each requirement, and to determine the overall criticality by combining these functions using the maximum. This is reasonable since the criticality function should reflect the more critical property for every simulation run. Note that this is a crucial modeling decision. Alternatively, a weighted sum of the functional components could be considered to reflect both properties in the criticality values. However, for the present case study no such importance weighting of the requirements is given. Apart from that, a weighted sum of criticality components may lead to the fact that ---depending on the chosen weighting--- a run for which one of the properties is highly critical while the other property is absolutely uncritical, results in a fairly low criticality value. Clearly, this might be disadvantageous when parameter configurations are to be detected that lead to either a high charging time or a high battery temperature. 

For both properties, the (single) auxiliary criticality is modeled to depend linearly on the properties regarding the requirements. That is, the higher the charging time (and battery temperature, respectively), the higher should be the single criticality. Thus, a linear shape for both function components $\kappa_\text{time}$ and $\kappa_\text{temp}$ is adequate. Furthermore, the function components should also yield values bounded between 0 and 1, each corresponding to the minimum and maximum admissible values, respectively, for $t_\text{charge}=t_\text{charge}(x)$ and $T_\text{bat}=T_\text{bat}(x)$. By this and Eq.~\eqref{eq:fatal-values}, we obtain the \textit{auxiliary criticality functions} for the charging time
$$\kappa_\text{time}(t_\text{charge}) = \min \left\lbrace \frac{t_\text{charge}-t_\text{min}}{t_\text{charge,fatal}-t_\text{min}} ,1 \right\rbrace = \min \left\lbrace \frac{t_\text{charge}}{9\ \text{h}} ,1 \right\rbrace\,,$$
where $t_\text{min} = 0~\text{h}$ is the minimal charging time, and for the battery temperature 
$$\kappa_\text{temp}(T_\text{bat}) = \min \left\lbrace \frac{T_\text{bat}-T_\text{min}}{T_\text{bat,fatal}-T_\text{min}} ,1 \right\rbrace = \min \left\lbrace \frac{T_\text{bat}+5\ ^\circ\text{C}}{{68.75\ }^{\circ}\text{C}},1\right\rbrace\,,$$
where $t_\text{min} = -5~^\circ\text{C}$ represents the minimal battery temperature that can be reached.

Finally, the overall \textit{criticality function} $\kappa$ is defined as the maximum of both components according to the considerations above, i.e., for any parameter point (or configuration) $x\in X$: 
\begin{equation}\label{eq:crit-function}
\kappa(x) = \max \left\lbrace \kappa_\text{time}(t_\text{charge}(x)), \kappa_\text{temp}(T_\text{bat}(x)) \right\rbrace\,.
\end{equation}

\subsection{Applicability}\label{subsec:applicability}
In order to show that the methodology described in Section~\ref{sec:methodology} is applicable and effective, it must be demonstrated that the assumptions of the respective algorithms applied to the present use case are fulfilled. Certainly, this is only necessary for the optimistic optimization algorithms to be applied (\texttt{HOO}, \texttt{POO}, \texttt{DOO}, \texttt{SOO}) since simple Monte Carlo simulation does not require any assumptions.\footnote{The only requirement is that the objective function (criticality function) can be evaluated pointwise, i.e. for every parameter point we have to be able to execute the corresponding simulation and to evaluate the (unique) criticality value of the run.} The aforementioned assumptions for optimistic optimization can be classified into two categories: those that concern the objective function, i.e. the \textit{criticality function}, and those that concern the \textit{partitioning} of the parameter space in the course of the algorithm (node splitting). In the following, both categories will be briefly examined to provide evidence for the existence of the assumptions.

Most of the assumptions regarding the criticality function are implied to hold if we can reason that the criticality function defined above can be regarded as Lipschitz continuous w.r.t. a suitable metric. In our case, we can presume that this is fulfilled for the following reason. First, we operate on simple spaces, namely the parameter space (domain) $X$ being a bounded, two-dimensional product space of real parameters, and the co-domain $\kappa(X)$ which is a (bounded) interval of real criticality values. Hence, the Lipschitz continuity can be regarded w.r.t. the standard Euclidean metric in our case. Intuitively, this means that the criticality function may not decrease too fast around its global optima (which unknown a priori). If we now assume that this is not fulfilled, i.e. if the criticality function would decrease too fast, then it would exhibit at least one global optimum in shape of a sharp peak (or even a single point in case of discontinuity). In particular, it would be impossible to derive the function’s behavior locally around that peak. That, however, contradicts the nature of the SUT as this would mean that small modifications in the parameter space (charging current and/or ambient temperature) can result in large variations or even jumps of charging time or temperature of the battery, according to the above definition of the criticality. Certainly, we can assume the SUT to be not that sensitive w.r.t. the aforementioned parameter inputs. Apart from that, assuming the criticality function to exhibit such steep peaks would make a \textit{sampling-based} approach inadequate when those peaks are to be found -- this would somehow correspond to finding a needle in a haystack.

The only aspect regarding the Lipschitz continuity that we cannot easily determine beforehand is the exact degree of smoothness of the criticality function, i.e. its Lipschitz constant in a broader sense. Thus, we have to test several instances with varying smoothness parameters for every optimistic optimization algorithm (see Sec.~\ref{subsec:simulation}) and evaluate how well-suited these are each for the purpose of finding critical events in our case. 

Regarding the partitioning of the search space, we employ a rather standard variant to split the subregions. More precisely, we use a binary tree of nodes (i.e. $K=2$), transform the parameter space into a normalized hypercube $[0,1]\times[0,1]$, and split the subdomain associated with a node in the middle along the longest side of the cell. That is, we identify $X=P_{0,1}$, and obtain $2^h$ cells for depth $h\ge0$ of same shape and size. This can be considered as splitting the two-dimensional hypercube regularly into smaller hypercubes, which shrink exactly at a geometric rate, namely with $\nu_1\rho^h$ (cf.~\cite{Bubeck2011}). In this way, the cells have bounded diameters on the one hand (bounded by $\nu_1\rho^h$), and are well-shaped on the other hand, i.e. it must be possible to embed pairwise non-overlapping, open balls into every cell whose diameters are bounded by some $\nu_2\rho^h$. Intuitively, the latter property requires that the cells will not collapse or have too sharp angles when $h$ increases. Both these properties of the cells are required for \texttt{HOO}/\texttt{POO} (cf.~\cite[Assumption 1]{Bubeck2011}) and \texttt{DOO}/\texttt{SOO} (cf.~\cite[Assumptions 3-4]{munos}) if in the latter case we identify the required semi-metric $l$ with the two-dimensional Euclidean metric and the decreasing sequence $\delta(h)$ with $\nu_1\rho^h$. In the course of the present case study, we choose to make these identifications mainly for reasons of comparability of the different algorithms, but also for the fact that for these choices, the fulfillment of the respective assumptions is given according to \cite[Example 1]{Bubeck2011}. However, as the smoothness of the criticality function and the just mentioned cell partitioning are related, again we do not know the (semi-)metric $l$ and the smoothness parameters $\nu_1$ and $\rho$ exactly for the present case so that we have to test and evaluate several combinations for all proposed algorithms.

\subsection{Implementation}\label{subsec:implementation}
The different models of the overall system under test (\textit{Battery}, \textit{Charging Approval}, \textit{Charging Management}, \textit{Charging Station} and \textit{Criticality Monitor}) have been developed in \textsc{Simulink} \cite{Simulink} which is a graphical tool for designing and implementing dynamical systems using \textsc{Matlab} \cite{MATLAB} as programming language. Both \textsc{Matlab} and \textsc{Simulink} are developed by MathWorks. For the implementation, \textsc{Matlab} 2017a and \textsc{Simulink} Version 8.9 have been used. The developed models have been exported as FMUs Version 2.0 (Functional-Mock Up Units) using the \textsc{FMI Kit for Simulink} Version 2.4.2 from Dassault Systems \cite{FMIKit}. The FMI Standard is a standardized interface for simulation models developed by the Modelica Association. A simulation model that implements this standard is called an “FMU” \cite{Modelica2014}. The co-simulation has been performed using \textsc{MasterSim} Version 0.8.2 from Bauklimatik Dresden \cite{MasterSim} as co-simulation engine. \textsc{MasterSim} was configured to use the Gauss-Seidel algorithm \cite{nicolai2018co} for the communication between the different simulation models with a communication step size of $t_\text{step}=1$ s. 

Additionally, the OFFIS in-house prototype tool \textit{OFFIS StreetTools} developed within the scope of the ENABLE-S3 project\footnote{\url{https://enable-s3.eu/}} has been reused. The \textit{StreetTools} operate as the test manager in our setup, executing the algorithms and creating new test cases based on previous observations. These test cases are passed to \textsc{MasterSim} which performs the actual simulations.

\subsection{Simulation}\label{subsec:simulation}
Each simulation starts with an empty battery\footnote{This is indeed the worst, probably rather unlikely case in the real world. However, let us recall that we consider \textit{models} here for which comparability is the only guiding principle.} ($SoC=0$) and normal temperature condition for the battery ($T_\text{bat}=20^\circ$C). As each simulation shows a single charging process of the battery, the simulation will be stopped when either the battery is fully charged ($SoC\ge0.95$) or when the charging process is still not finished after nine hours of simulated time (leading to the maximum criticality value of 1 as it was not possible to finish the charging process). After the simulation ends the highest measured criticality value during that simulation run is reported back to the test manager and a new simulation run is instantiated according to the algorithm chosen beforehand. For each of the algorithms the simulation budget was set to $N_\text{budget}=4000$ simulations.\\

\begin{figure*}[h!]
	\centering
	\includegraphics[width=\textwidth]{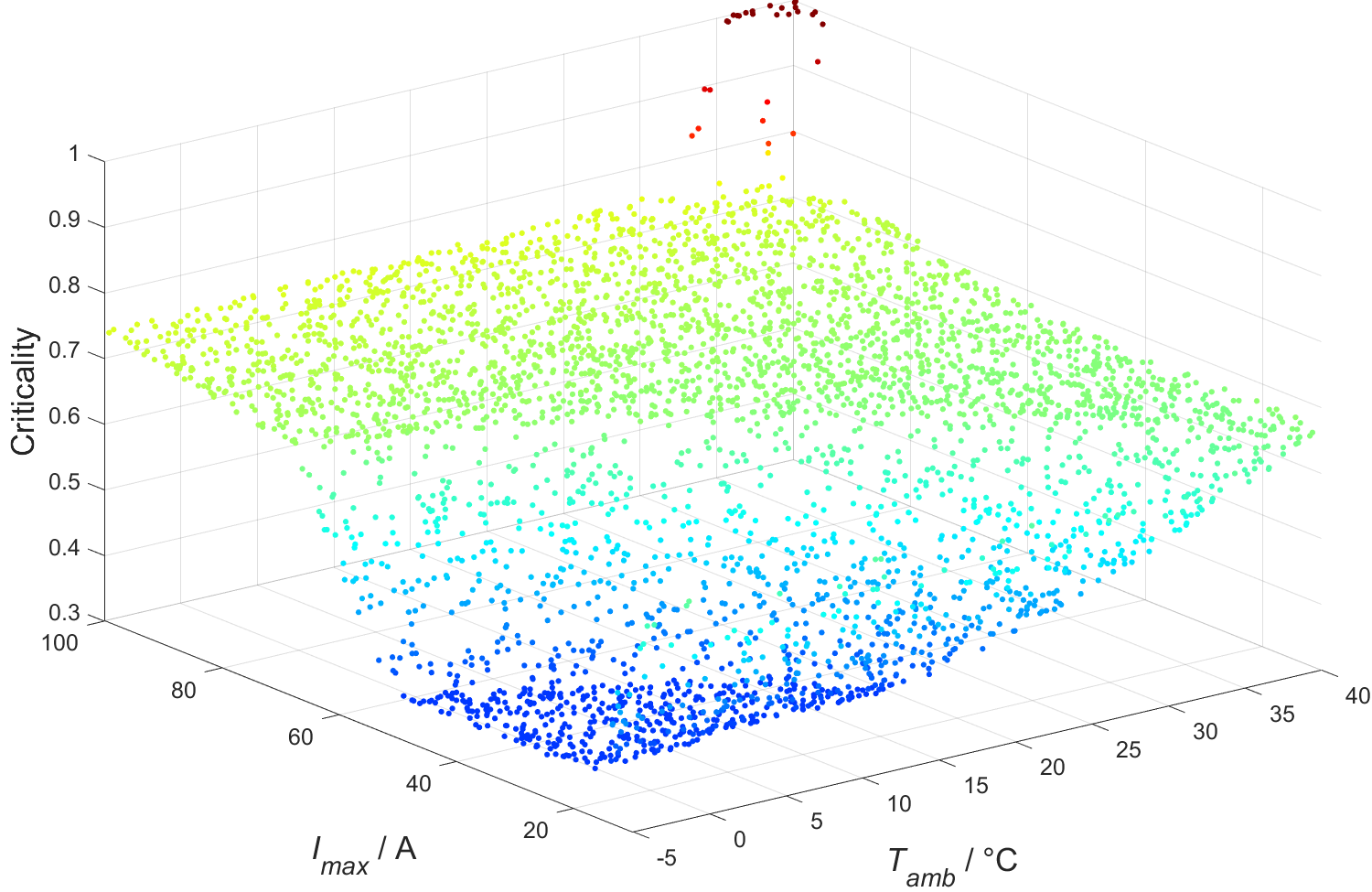}
	\caption{Overview of the results using Monte Carlo simulation. \textcolor{BrickRed}{Red} marked points indicate a criticality higher than the given threshold $c_\kappa=0.8$.}\label{fig:plot-MC1}
\end{figure*}

\begin{figure*}[t!]
	\centering
	\includegraphics[width=\textwidth]{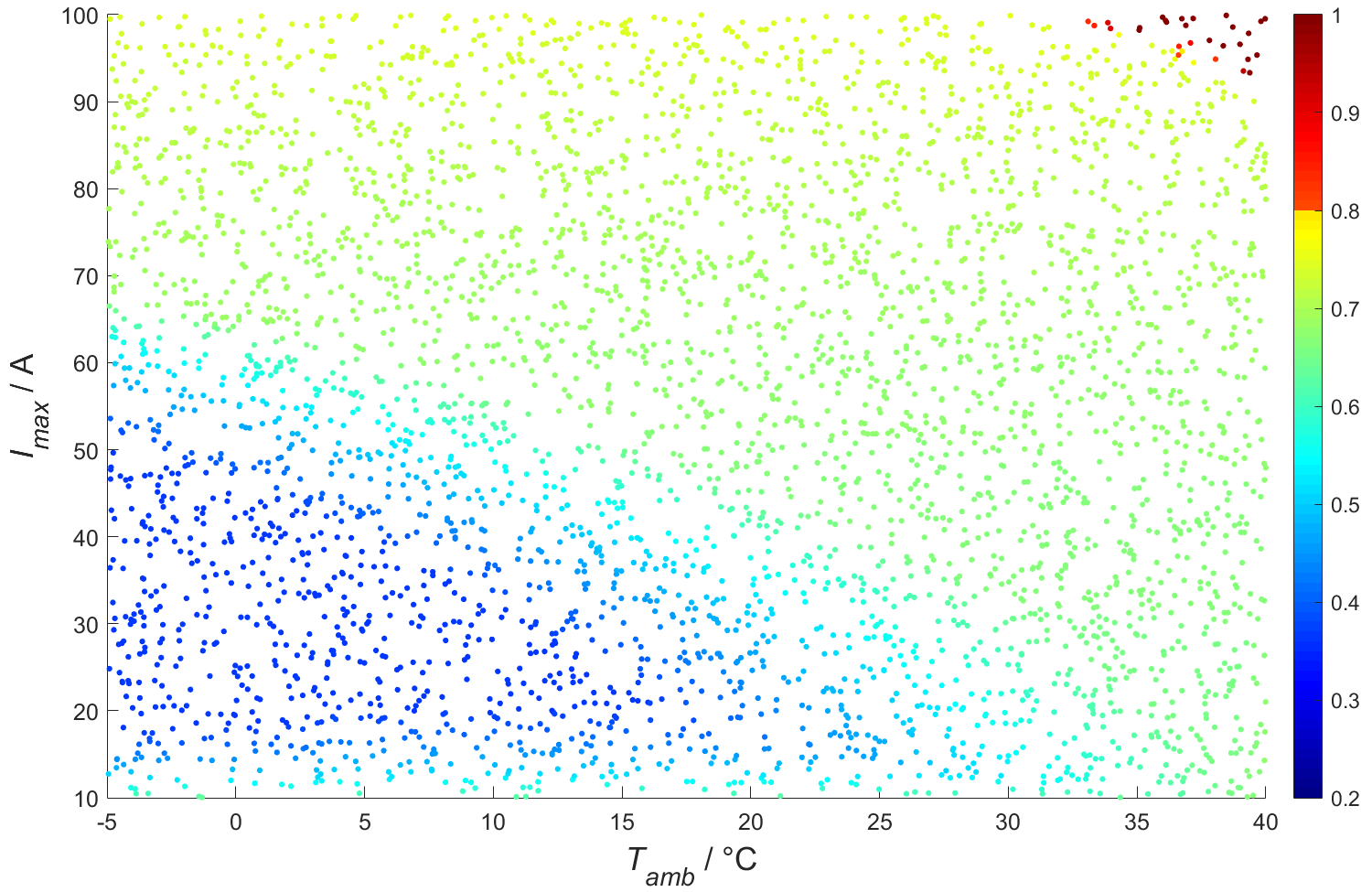}
	\caption{Heat map of the Monte Carlo simulation results from Figure~\ref{fig:plot-MC1}. \textcolor{BrickRed}{Red} marked points indicate a criticality higher than the given threshold $c_\kappa=0.8$.}\label{fig:plot-MC2}
\end{figure*}

\hspace*{-0.55cm}\underline{Monte Carlo Simulation}\\
As a baseline for comparison with the other algorithms, we applied simple Monte Carlo simulation (uniform sampling) to the use case. The results of our simulations are visualized in Figures~\ref{fig:plot-MC1} and~\ref{fig:plot-MC2} wherein the measured criticality is plotted versus all tested parameter configurations. Red dots indicate a critical test run for which the criticality exceeds the threshold of $0.8$ (i.e., a critical event) while all other colors indicate harmless behavior. 

We can group the parameter space primarily in three different regions. The first region (colored in deep blue) is where both the temperature and the current are quite low. There we measure a very low criticality of around $0.36$. The criticality rises fast if the available maximum current either decreases too much (around $I_\text{max}<17$) as this increases the charging time heavily, or if the current increases too much as this leads to an increased heating of the battery and thus to higher temperatures and a higher criticality. In the second region (colored in green and yellow) the criticality is quite flat as here the temperature is controlled by the Battery Management System that keeps decreasing temporarily the current to keep the temperature from rising too much. This “safety mechanism” does not work in the third region, the critical region, where additionally also the ambient temperature is very high so that no additional cooling from the outside is provided. As it can be expected, those extreme parameter configurations challenge the Battery Management System and the battery will either shortly overheat or the charging process will be (temporarily) interrupted leading to very high charging times.

This experiment with Monte Carlo has been carried out five times. We measured the following numbers of rare critical events: $28$ (these simulations are depicted in Figs.~\ref{fig:plot-MC1} and~\ref{fig:plot-MC2}), 31, 24, 32 and 32 again. In the average we observed $29.40 \pm 3.44$ (standard deviation) critical events using the Monte Carlo method leading to a critical event ratio of around $0.7\%$ which means that $99.3\%$ of the simulation budget is spent on uncritical cases. Also due to the low number of measuring points in the critical area we do not get a clear picture of the transition zone between critical and uncritical cases. \\

\hspace*{-0.55cm}\underline{HOO}\\
As outlined earlier, the instantiation of the \texttt{HOO} algorithm requires to determine the hyperparameters $\nu_1$ and $\rho$ that depend on the smoothness of the criticality function. We choose $\nu_1=1.0$ which is a suitable value as the criticality function is modeled to take values between $0.0$ and $1.0$, which makes $1.0$ the maximum possible difference between the criticality values for two arbitrary points in $X$, and thus it is a valid upper bound for the constant growth factor $\nu_1$. The choice of the second hyperparameter $\rho$ is more difficult as the smoothness of the criticality function $\kappa(x)$ with $x= (T_\text{amb}, I_\text{max}$) is unknown a priori. For that reason, we created ten different instantiations between $\rho=0.1$ and $\rho=0.99$. 


\begin{table}
	\centering
	\begin{tabular}{| l | l |}
		\hline
		\; \textbf{Algorithm} \; & \; \textbf{Critical events} \;\\
		\hline 
		\; \texttt{HOO} ($\nu_1 = 1.0, \rho = 0.1$) \; & \; $2064$  \\
		\hline 
		\; \texttt{HOO} ($\nu_1 = 1.0, \rho = 0.2$) \; & \; $2041$  \\
		\hline 
		\; \texttt{HOO} ($\nu_1 = 1.0, \rho = 0.3$) \; & \; $1930$  \\
		\hline 
		\; \texttt{HOO} ($\nu_1 = 1.0, \rho = 0.4$) \; & \; $1874$  \\
		\hline
		\; \texttt{HOO} ($\nu_1 = 1.0, \rho = 0.5$) \; & \; $1576$  \\
		\hline  
		\; \texttt{HOO} ($\nu_1 = 1.0, \rho = 0.6$) \; & \; $1008$ \\
		\hline 
		\; \texttt{HOO} ($\nu_1 = 1.0, \rho = 0.7$) \; & \; $579$  \\
		\hline
		\; \texttt{HOO} ($\nu_1 = 1.0, \rho = 0.8$) \; & \; $422$  \\
		\hline  
		\; \texttt{HOO} ($\nu_1 = 1.0, \rho = 0.9$) \; & \; $633$  \\
		\hline
		\; \texttt{HOO} ($\nu_1 = 1.0, \rho = 0.99$) \; & \; $2132$  \\
		\hline  
		\; Monte Carlo (average) \; & \; $29.40$  \\
		\hline 
	\end{tabular}
	\vspace{0.3cm}
	\caption{Number of rare critical events after $N = 4000$ simulations for different hyperparameter configurations.}\label{tab:HOO-results}
\end{table}

\begin{figure*}[h!]
	\centering
	\includegraphics[width=\textwidth]{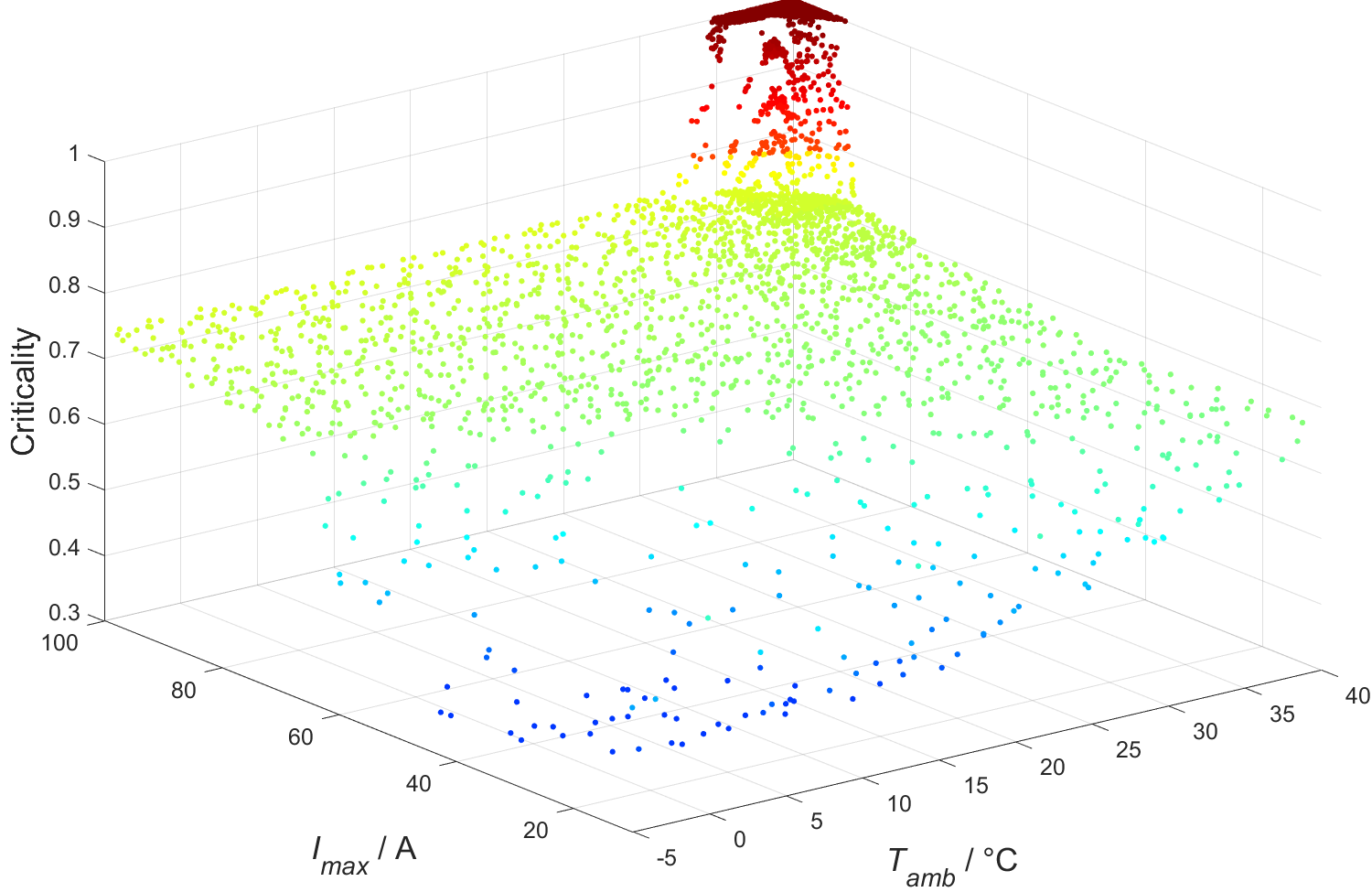}
	\caption{Overview of the results applying the \texttt{HOO} algorithm with $\nu_1=1.0$ and $\rho=0.3$. \textcolor{BrickRed}{Red} marked points indicate a criticality higher than the given threshold $c_\kappa=0.8$.}\label{fig:plot-HOO1}
\end{figure*}

\begin{figure*}[h!]
	\centering
	\includegraphics[width=\textwidth]{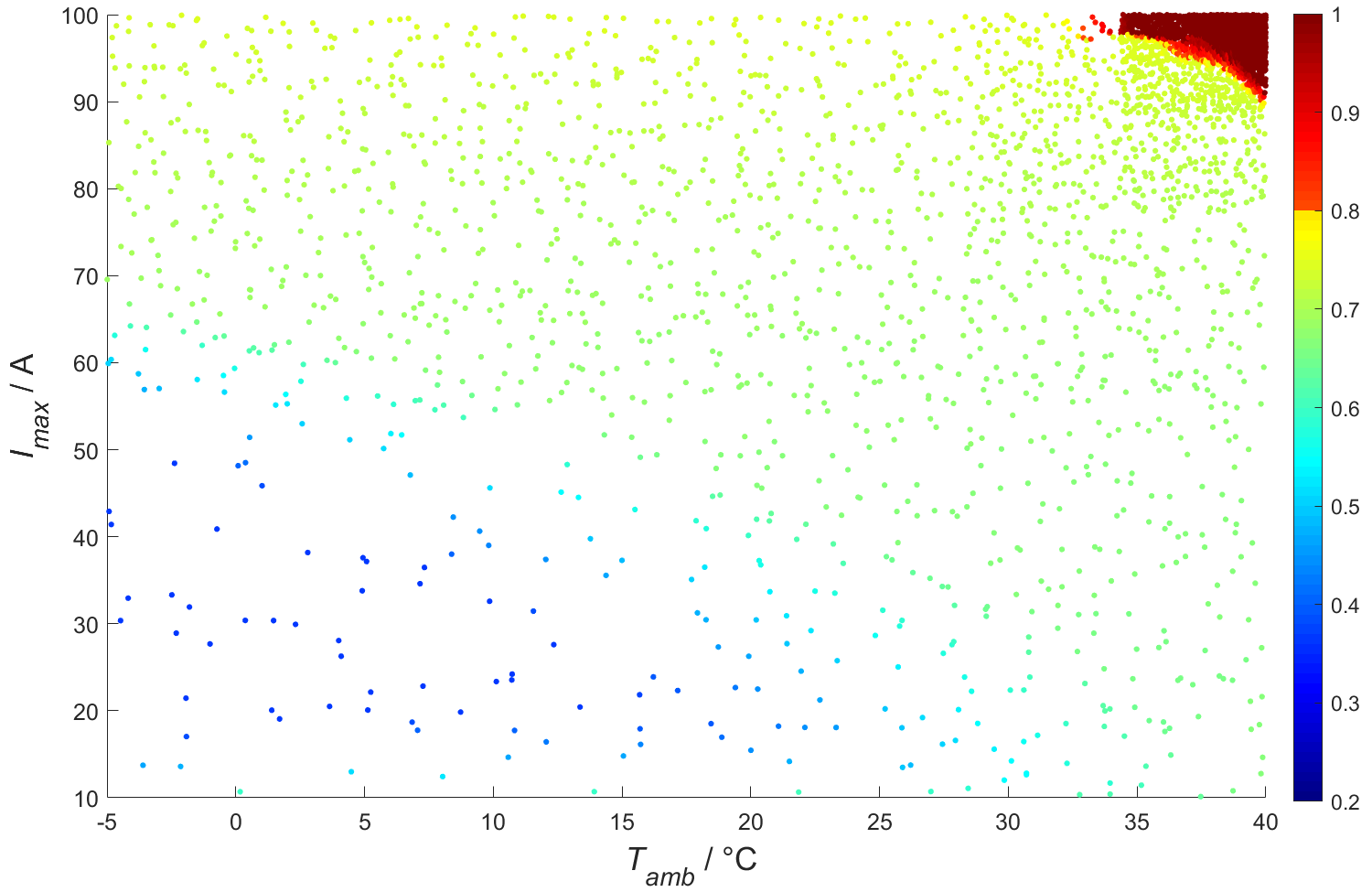}
	\caption{Heat map of the \texttt{HOO} algorithm results with $\nu_1=1.0$ and $\rho=0.3$ from Figure~\ref{fig:plot-HOO1}. \textcolor{BrickRed}{Red} marked points indicate a criticality higher than the given threshold $c_\kappa=0.8$.}\label{fig:plot-HOO2}
\end{figure*}

The results can be found in Table~\ref{tab:HOO-results}. The main result that can be taken from this is that for all tested hyperparameter combinations \texttt{HOO} found much more rare events than Monte Carlo, thus showing the strength of this approach. An exemplary graphical depiction of one of the instances ($\rho=0.3$) can be found in Figures~\ref{fig:plot-HOO1} and~\ref{fig:plot-HOO2}. In comparison to the results from the Monte Carlo simulation, we get a notably more precise depiction of the critical area in the top right corner of the parameter space.

Compared against each other, the different \texttt{HOO} instances also performed very differently (again in the sense of number of critical events found). The best performing instance ($\rho=0.99$) had more than five times as many critical events as the worst instance ($\rho=0.8$) which confirms the importance of well-defined hyperparameters. Overall the algorithm showed its best performance on this use-case with either low values for $\rho$ ($\rho = 0.1, ..., 0.4$) or with a very high $\rho$ value ($\rho=0.99$). (Note, that $\rho = 0.9$ performed, compared to the other instances, not so well, so we can assume that the number of critical events will raise rapidly when $\rho > 0.9$ on this use-case).

\begin{figure*}[h!]
	\centering
	\includegraphics[width=\textwidth]{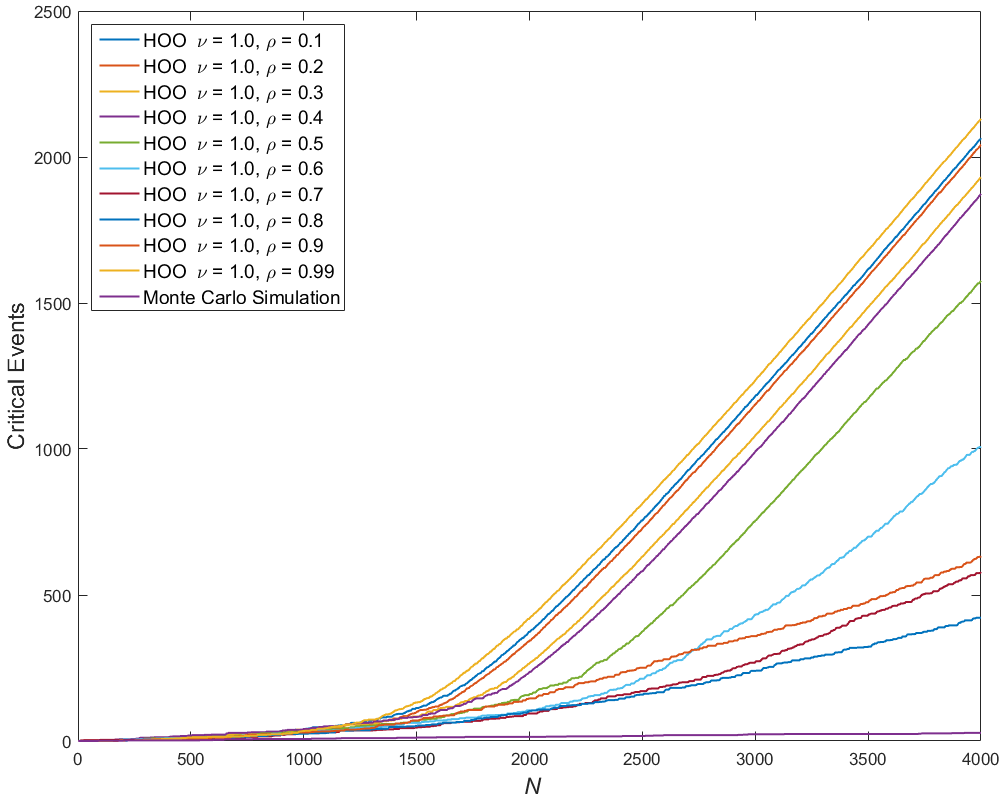}
	\caption{Accumulated number of detected rare events for different \texttt{HOO} instances (compared to simple Monte Carlo) after $N$ simulations.}\label{fig:graphs-HOO-vs-MC}
\end{figure*}

More details on the performance of the \texttt{HOO} algorithm can be seen in Figure~\ref{fig:graphs-HOO-vs-MC}. Here, we compare the number of found rare events in dependence of the number of simulations $N$ being performed. An important lesson that we can take from the depiction is that a sufficient number of simulation runs is needed for the \texttt{HOO} algorithm to perform well. For instance, for up to $N = 1300$ the different instances of \texttt{HOO} all performed roughly equally well. One possible explanation for this behavior is that for a low number of simulations the uncertainty is relatively high as many nodes have not been visited very often, i.e., the uncertainty term $\sqrt{2 \ln(N)/T_{h,i}}$ in the $U$-values (see Eq.~\eqref{eq:U-value}) is comparatively large and thus overshadows the smoothness term $\nu_1 \rho^h$. 

\begin{figure*}[h!]
	\centering
	\includegraphics[width=\textwidth]{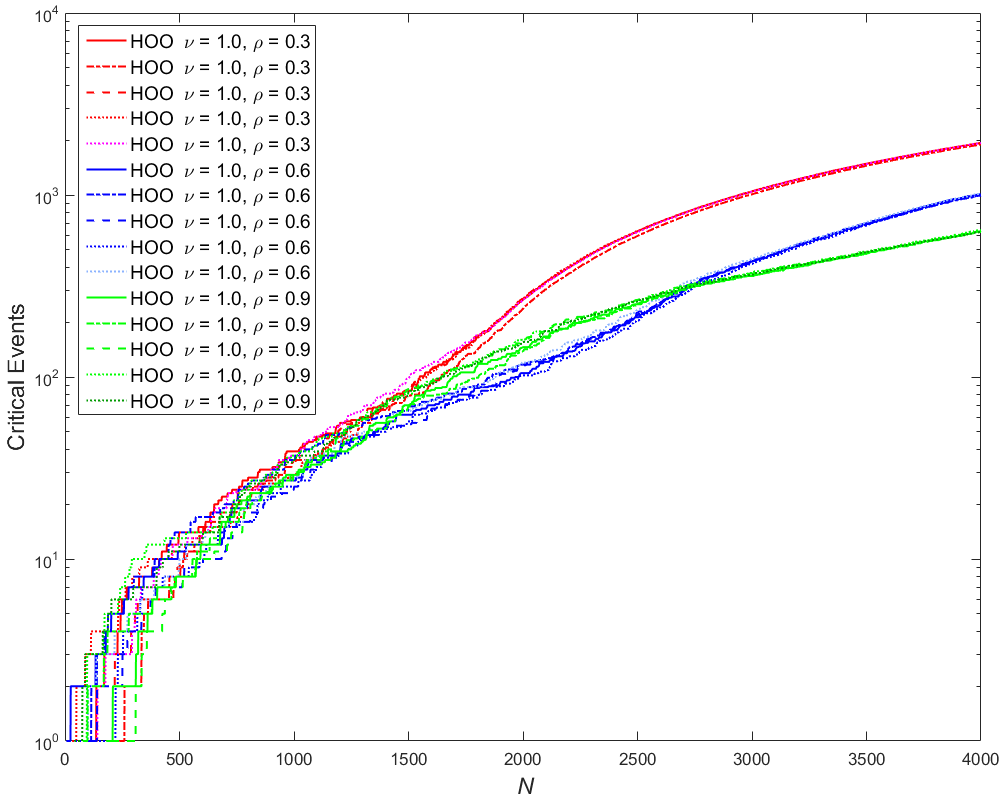}
	\caption{Accumulated number of rare events for different hyperparameter choices and runs of the \texttt{HOO} algorithm on a semi-logarithmic scale.
	}\label{fig:graphs-HOOs}
\end{figure*}

Another important aspect that can be taken from this is how different start conditions affect the performance of \texttt{HOO}. As the \texttt{HOO} always samples a random point in the chosen subdomain its performance could depend on the chance of sampling “good” points in the beginning. With an increasing number of simulations this should balance out which our results also indicate (see Figure~\ref{fig:graphs-HOOs}). There we can see five different runs of the \texttt{HOO} algorithm for three different hyperparameter configurations. We choosed $\rho = 0.3$, $\rho = 0.6$ and $\rho = 0.9$ as these are three very different equidistant parameter configurations that performed very differently (see Table \ref{tab:HOO-results}). The red curves are different runs for the smoothness parameters $(\nu_1,\rho)=(1.0, 0.3)$. As shown in the plot, with an increasing number of simulations $N$ the red curves converge to the same curve indicating that 4000 simulations are enough to balance out the different start conditions. The same is also true for the other tested instances $(1.0, 0.6)$ (blue) and $(1.0, 0.9)$ (green). We find the following averages and their standard deviations over five runs of each instance: $1925.40 \pm 16.09$ rare events for the red curves, $1004.60 \pm 10.01$ (blue curves), and $631.60 \pm 6.19$ (green). Overall, we can also assume the different start conditions will not affect the simulation results long term. This has been exemplarily tested with only five instances per configuration and only on the three pictured configurations due to simulation time restrictions.

Additionally note that, once sufficient simulation data is available (approx. $N\ge2750$), the number of detected rare events increases linearly with the number of simulations being performed (taking into account the logarithmic scaling of the y-axis). This indicates again the strength of the \texttt{HOO} algorithm w.r.t. detecting rare events efficiently.\\

\hspace*{-0.55cm}\underline{POO}\\
The \texttt{POO} algorithm is specifically designed for the case that the smoothness of the criticality function is unknown, i.e., if the smoothness parameters $\nu_1$ and $\rho$ are not available beforehand. The hyperparameters of \texttt{POO}, $\nu_\text{max}$ and $\rho_\text{max}$, define the payoff between exploration and exploitation by influencing the numbers of instances being created. In our case, we choosed the same values for $\nu_{max}$ and $\rho_{max}$ as for $\nu$ and $\rho$ for the \texttt{HOO} algorithm (see above). Please note, that with decreasing $\rho_{max}$ less \texttt{HOO} instances will be created which will make the algorithm more similiar to the \texttt{HOO} algorithm. 

In our setup we enabled the communication between the different \texttt{HOO} instances within the \texttt{POO} algorithm. That is, whenever a \texttt{HOO} instance requests a simulation of a node, a “look-up” will be performed instead of instantly simulating for the corresponding node. This means that all \texttt{HOO} instances will be checked if that node has already been simulated by another instance. In that case, the previous result will be used which leads to saving simulation budget. This is an improvement that Grill et al.~suggested as well (cf.~\cite[Appendix D]{Grill2015}).

\begin{table}
	\centering
	\begin{tabular}{| p{0.4\textwidth} | p{0.25\textwidth} | p{0.32\textwidth} |}
		\hline
		\,\textbf{Algorithm} \; & \,\textbf{Number of \texttt{HOO} \newline \;\, instances created} \; & \,\textbf{Critical events} \;\\
		\hline 
		\; \texttt{POO} ($\nu_{\text{max}} = 1.0, \rho_\text{max} = 0.1$) \; & \; $1$ \; & \; $2023$  \\
		\hline 
		\; \texttt{POO} ($\nu_{\text{max}} = 1.0, \rho_\text{max} = 0.2$) \; & \; $2$ \; & \; $1774$  \\
		\hline 
		\; \texttt{POO} ($\nu_{\text{max}} = 1.0, \rho_\text{max} = 0.3$) \; & \; $2$ \; & \; $1676$  \\
		\hline 
		\; \texttt{POO} ($\nu_{\text{max}} = 1.0, \rho_\text{max} = 0.4$) \; & \; $4$ \; & \; $1255$  \\
		\hline 
		\; \texttt{POO} ($\nu_{\text{max}} = 1.0, \rho_\text{max} = 0.5$) \; & \; $4$ \; & \; $980$  \\
		\hline
		\; \texttt{POO} ($\nu_{\text{max}} = 1.0, \rho_\text{max} = 0.6$) \; & \; $8$ \; & \; $905$  \\
		\hline 
		\; \texttt{POO} ($\nu_{\text{max}} = 1.0, \rho_\text{max} = 0.7$) \; & \; $8$ \; & \; $787$  \\
		\hline 
		\; \texttt{POO} ($\nu_{\text{max}} = 1.0, \rho_\text{max} = 0.8$) \; & \; $16$ \; & \; $811$  \\
		\hline 
		\; \texttt{POO} ($\nu_{\text{max}} = 1.0, \rho_\text{max} = 0.9$) \; & \; $32$ \; & \; $801$  \\
		\hline 
		\; \texttt{POO} ($\nu_{\text{max}} = 1.0, \rho_\text{max} = 0.99$) \; & \; $512$ \; & \; $1036$  \\
		\hline 
	\end{tabular}
	\vspace{0.3cm}
	\caption{Number of rare critical events after $N = 4000$ simulations for different hyperparameter configurations.}\label{tab:POO-results}
\end{table}

\begin{figure*}[h!]
	\centering
	\includegraphics[width=0.95\textwidth]{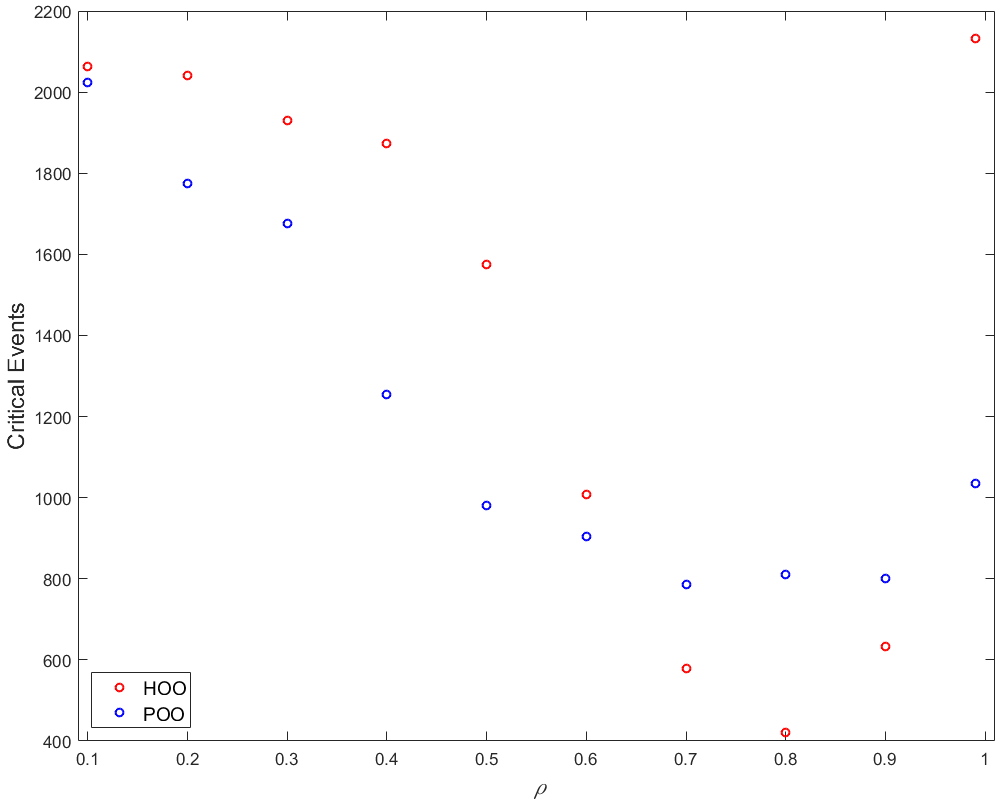}
	\caption{Number of found critical events for different  choices of the hyperparameter $\rho$ for the \texttt{HOO} and \texttt{POO} algorithm.
	}\label{fig:graphs-HOOandPOOs}
\end{figure*}

The results can be found in Table ~\ref{tab:POO-results}. A graphical depiction together with the \texttt{HOO} results can be found in Figure \ref{fig:graphs-HOOandPOOs}. As we can see there \texttt{HOO} yielded more critical events than \texttt{POO} for $\rho \in \{0.1, 0.2, 0.3, 0.4, 0.5, 0.6, 0.99\}$ and vice versa for $\rho \in \{0.7, 0.8, 0.9\}$. Overall, the behavior of both algorithms in Figure \ref{fig:graphs-HOOandPOOs} for the different $\rho$-values is quite similiar but the \texttt{HOO} algorithm is more extreme in the sense that it performs better for well-suited $\rho$-values and worse for inadequate $\rho$-values. This shows less dependence on the hyperparameters for the \texttt{POO} algorithm which was expected as \texttt{POO} is designed particularly for the case of an unknown smoothness of the criticality function. With increasing $\rho_{\text{max}}$, more instances are tested leading to both more “good” and more “bad” instances which balance each other out to some extent. Overall, this means that the \texttt{POO} algorithm performs always worse than a \texttt{HOO} instance being optimally fitted to the problem. Fortunately, this loss is bounded and an estimate of the loss is given in the original publication~\cite{Grill2015}.

The \texttt{POO} algorithm can also be used to obtain a well-fitting hyperparameter pair for the \texttt{HOO} algorithm. For that purpose, we recommend to set $\rho_{\text{max}}$ rather high as this leads to a higher number of tested instances and hence to a high coverage of possible hyperparameter configurations. Especially when simulation results are shared between the instances, e.g., when look-ups are performed, it is much better to have a wide range of instances to get accurate information about the performance of the different hyperparameters.\\

\hspace*{-0.55cm}\underline{DOO}\\
Due to their nature of being multi-armed bandit algorithms, \texttt{HOO} and \texttt{POO} were primarily designed for non-deterministic systems for which the criticality values are potentially noisy as explained earlier. The SUT of the present case study is a deterministic system, hence there is a good chance that \texttt{DOO} performs better since it is potentially better suited for this purpose, although being more narrow regarding its range of application. 

\begin{figure*}[h!]
	\centering
	\includegraphics[width=\textwidth]{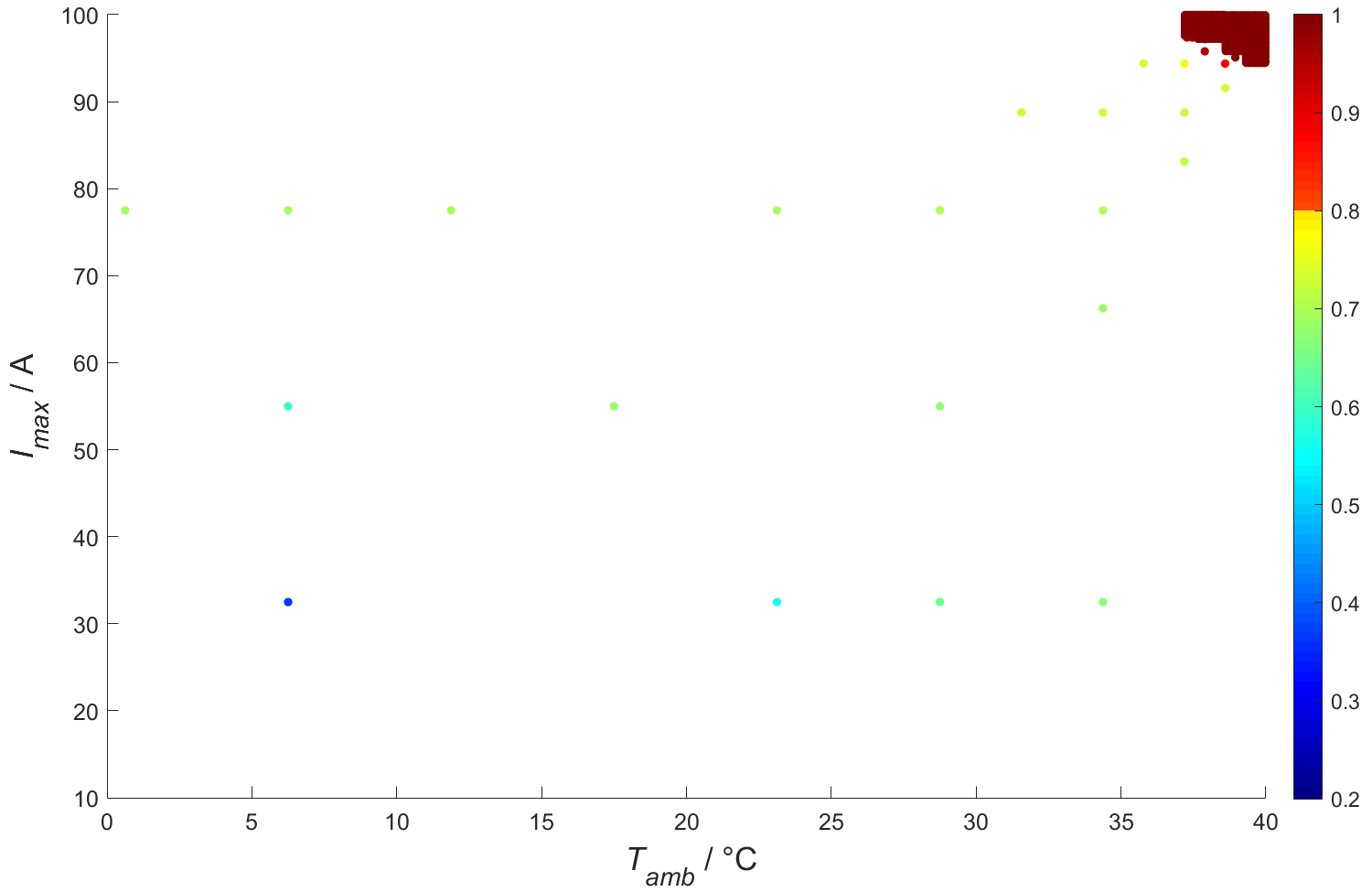}
	\caption{Heat map of the \texttt{DOO} algorithm results with $\delta(h) = \nu_1 \rho^h$, where $\nu_1=1.0$ and $\rho=0.3$. \textcolor{BrickRed}{Red} marked points indicate a criticality higher than the given threshold $c_\kappa=0.8$.}\label{fig:plot-DOO1}
\end{figure*}

\begin{figure*}[h!]
	\centering
	\includegraphics[width=\textwidth]{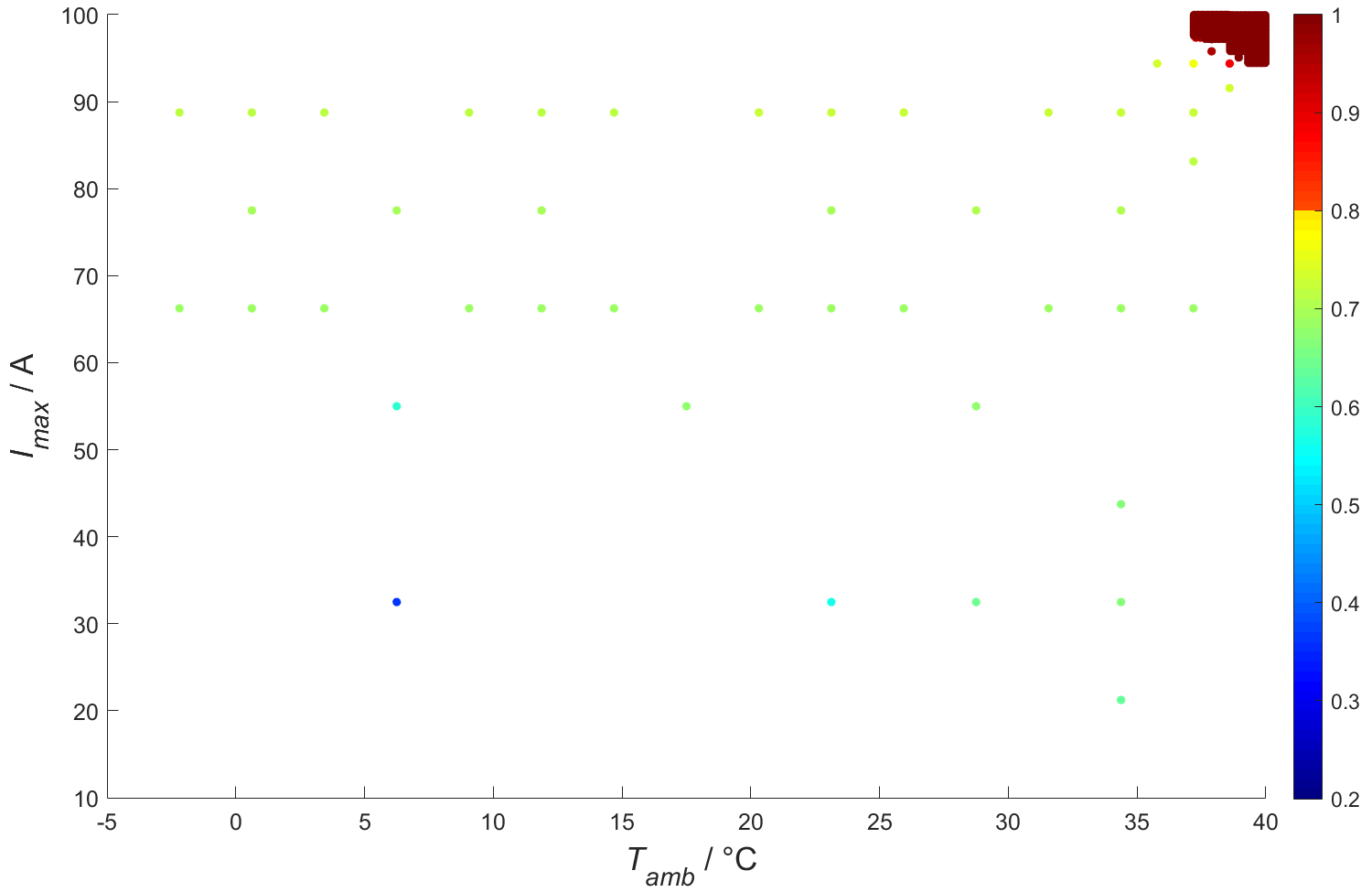}
	\caption{Heat map of the \texttt{DOO} algorithm results with $\delta(h) = \nu_1 \rho^h$, where $\nu_1=1.0$ and $\rho=0.6$. \textcolor{BrickRed}{Red} marked points indicate a criticality higher than the given threshold $c_\kappa=0.8$.}\label{fig:plot-DOO2}
\end{figure*}

\begin{figure*}[h!]
	\centering
	\includegraphics[width=\textwidth]{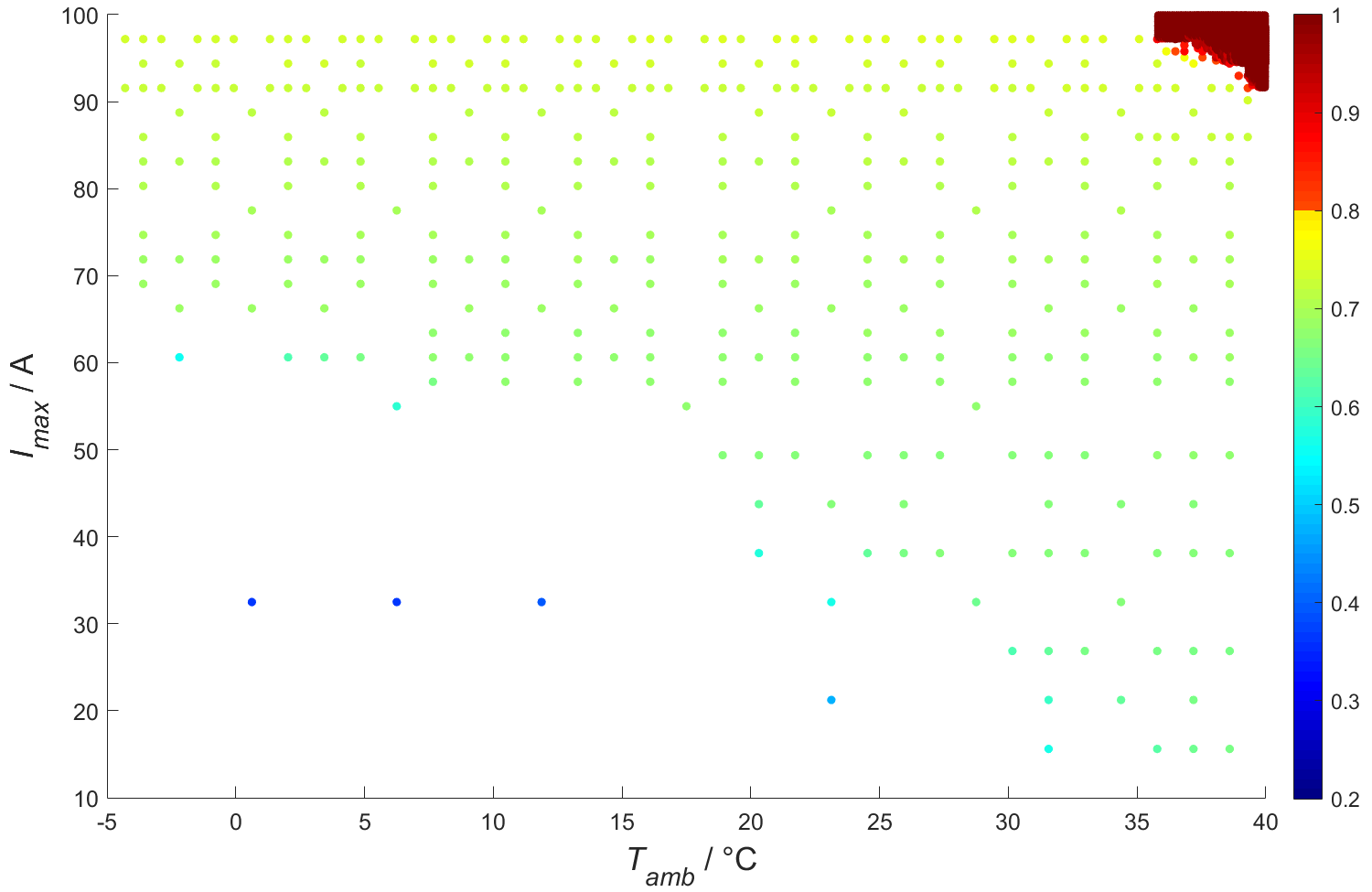}
	\caption{Heat map of the \texttt{DOO} algorithm results with $\delta(h) = \nu_1 \rho^h$, where $\nu_1=1.0$ and $\rho=0.9$. \textcolor{BrickRed}{Red} marked points indicate a criticality higher than the given threshold $c_\kappa=0.8$.}\label{fig:plot-DOO3}
\end{figure*}

As input argument we need a decreasing sequence $\delta(h)$, which we set as $\delta(h)=\nu_1 \rho^h$ analogously as for the \texttt{HOO} algorithm for the sake of comparability. We tested the algorithm with the same hyperparameters as for the \texttt{HOO} algorithm. The results can be taken from Table~\ref{tab:DOO-results} and Figures~\labelcref{fig:plot-DOO1,fig:plot-DOO2,fig:plot-DOO3}. Each hyperparameter configuration has been only simulated once as the \texttt{DOO} algorithm is fully deterministic.

\begin{table}[h!]
	\centering
	\begin{tabular}{| l | l |}
		\hline
		\; \textbf{Algorithm} \; & \; \textbf{Critical events} \;\\
		\hline 
		\; \texttt{DOO} ($\nu_1 = 1.0, \rho = 0.1$) \; & \; $3985$ \;  \\
		\hline 
		\; \texttt{DOO} ($\nu_1 = 1.0, \rho = 0.2$) \; & \; $3981$ \;  \\
		\hline 
		\; \texttt{DOO} ($\nu_1 = 1.0, \rho = 0.3$) \; & \; $3979$ \;  \\
		\hline 
		\; \texttt{DOO} ($\nu_1 = 1.0, \rho = 0.4$) \; & \; $3967$ \;  \\
		\hline 
		\; \texttt{DOO} ($\nu_1 = 1.0, \rho = 0.5$) \; & \; $3965$ \;  \\
		\hline 
		\; \texttt{DOO} ($\nu_1 = 1.0, \rho = 0.6$) \; & \; $3957$ \; \\
		\hline 
		\; \texttt{DOO} ($\nu_1 = 1.0, \rho = 0.7$) \; & \; $3937$ \;  \\
		\hline 
		\; \texttt{DOO} ($\nu_1 = 1.0, \rho = 0.8$) \; & \; $3933$ \;  \\
		\hline 
		\; \texttt{DOO} ($\nu_1 = 1.0, \rho = 0.9$) \; & \; $3641$ \; \\
		\hline 
		\; \texttt{DOO} ($\nu_1 = 1.0, \rho = 0.99$) \; & \; $3985$ \;  \\
		\hline 
	\end{tabular}
	\vspace{0.3cm}
	\caption{Number of rare critical events after $N = 4000$ simulations for different hyperparameter configurations.}\label{tab:DOO-results}
\end{table}

We can take from this that \texttt{DOO} is much more effective in sampling critical parameter points than the previously applied algorithms. For all tested parameter configurations more than 90\% of the simulation budget is spent in the critical area which is notably more effective compared to \texttt{HOO} and \texttt{POO}. As the main reason for this increased effectiveness we regard the fact that \texttt{DOO} does not need to take into account any extra uncertainty by noisy values which results in much tighter upper optimistic bounds compared to \texttt{HOO} and \texttt{POO} for the individual subregions. \footnote{Let us note that for an entirely “fair” comparison of the algorithms, we would have to align the way the algorithms work, e.g. by omitting the uncertainty term $\sqrt{2 \ln(N)/T_{h,i}}$ in Eq.~\eqref{eq:U-value} that accounts for the stochastic nature.}  

Additionally, the \texttt{DOO} algorithm shows to be also much more robust against badly chosen $\rho$-values for this specific use case as the simulations suggest. In terms of the number of detected critical events, with our worst instance $\rho=0.9$ we still obtain much better results than with \texttt{HOO} using the best instance ($\rho_{HOO}=0.99$) and are still close to the best \texttt{DOO} instances using $\rho=0.1$ and $\rho=0.99$, which further verifies the \texttt{DOO} algorithm to be a much more suitable algorithm for this type of problem. Actually all tested \texttt{DOO} instances with the exception of $\rho = 0.9$ differ only very slightly in the number of critical events detected.  

A drawback of using \texttt{DOO} for this specific use case is that it explored the parameter space very little for all tested hyperparameter configurations. The algorithm rather focused investigation on the upper right corner where the criticality takes values close to 1 but it did not sample the contiguous transition regions where the criticality raises up to 1. That is, we obtain almost no information regarding the actual size of the critical region and only sweep the region close to the maximum. As criticality values of $0.8$ and higher already indicate violated requirements, this can be especially problematic in the context of safety-critical systems for which the set of critical parameters is to be identified as precisely as possible. 

\begin{figure*}[h!]
	\centering
	\includegraphics[width=\textwidth]{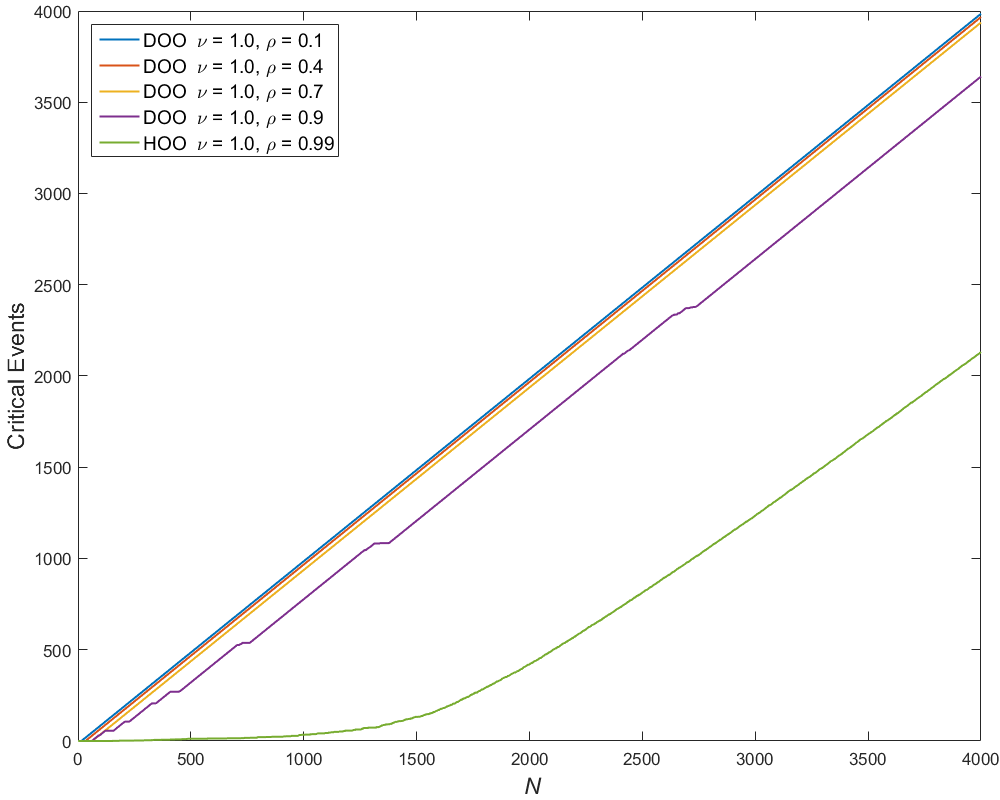}
	\caption{Accumulated number of detected rare events for different \texttt{DOO} instances (compared with \texttt{HOO}) after $N$ simulations.}\label{fig:graphs-DOO-vs-HOO}
\end{figure*}

One of the advantages of this method in the present context is that it needs much less simulation budget to actually guide the system into the critical regions compared to the \texttt{HOO} algorithm. This is depicted in Figure~\ref{fig:graphs-DOO-vs-HOO} where  four different \texttt{DOO} instances are compared to our best performing \texttt{HOO} instance. It can be seen that the number of found rare critical events for the \texttt{DOO} instances $\rho=0.1$, $\rho=0.4$ and $\rho=0.7$ raises approximately linearly with the increasing number of simulations. After very few explorative simulations in the beginning every simulated point is a critical event leading to a very high number of detected critical events. This is also true for the other not depicted \texttt{DOO} instances from Table \ref{tab:DOO-results}, which have been omitted from the figure for the sake of clarity. The only exception from this behavior is the $\rho = 0.9$ instance that has multiple comparatively longer exploration phases where the number of rare critical events is not or only slowly rising. Regardless, this instance still performed much better than the \texttt{HOO} instance, which needs much longer to gather data (which is caused by the fact that \texttt{HOO} is a stochastic algorithm trying to deal with eventual noisiness of the data). The very good performance of the \texttt{DOO} is also probably due to the “simple” nature of the parameter space for which the criticality increases with higher currents and temperatures.\\ 

\hspace*{-0.55cm}\underline{SOO}\\
Similar to the \texttt{POO} algorithm, the \texttt{SOO} algorithm can be used instead of \texttt{DOO} when the smoothness or rather the semi-metric $l$ is unknown for a deterministic system. As input argument a function $h_\text{max}(t)$ is needed that is defined in the number of node splittings $t$ that have been performed so far. In the original publication (cf.~\cite[Corollary 2]{munos}), the algorithm is analyzed for exponentially decreasing diameters with $\delta(h)=\nu \rho^h$ and $h_\text{max}(t)=t^\varepsilon$. For that reason we also used that definition of $h_\text{max}$ on this use case.\footnote{Recall that, in contrast to \texttt{DOO}, the application of \texttt{SOO} does not require to set the sequence $\delta(h)$.} 

\begin{figure*}[h!]
	\centering
	\includegraphics[width=\textwidth]{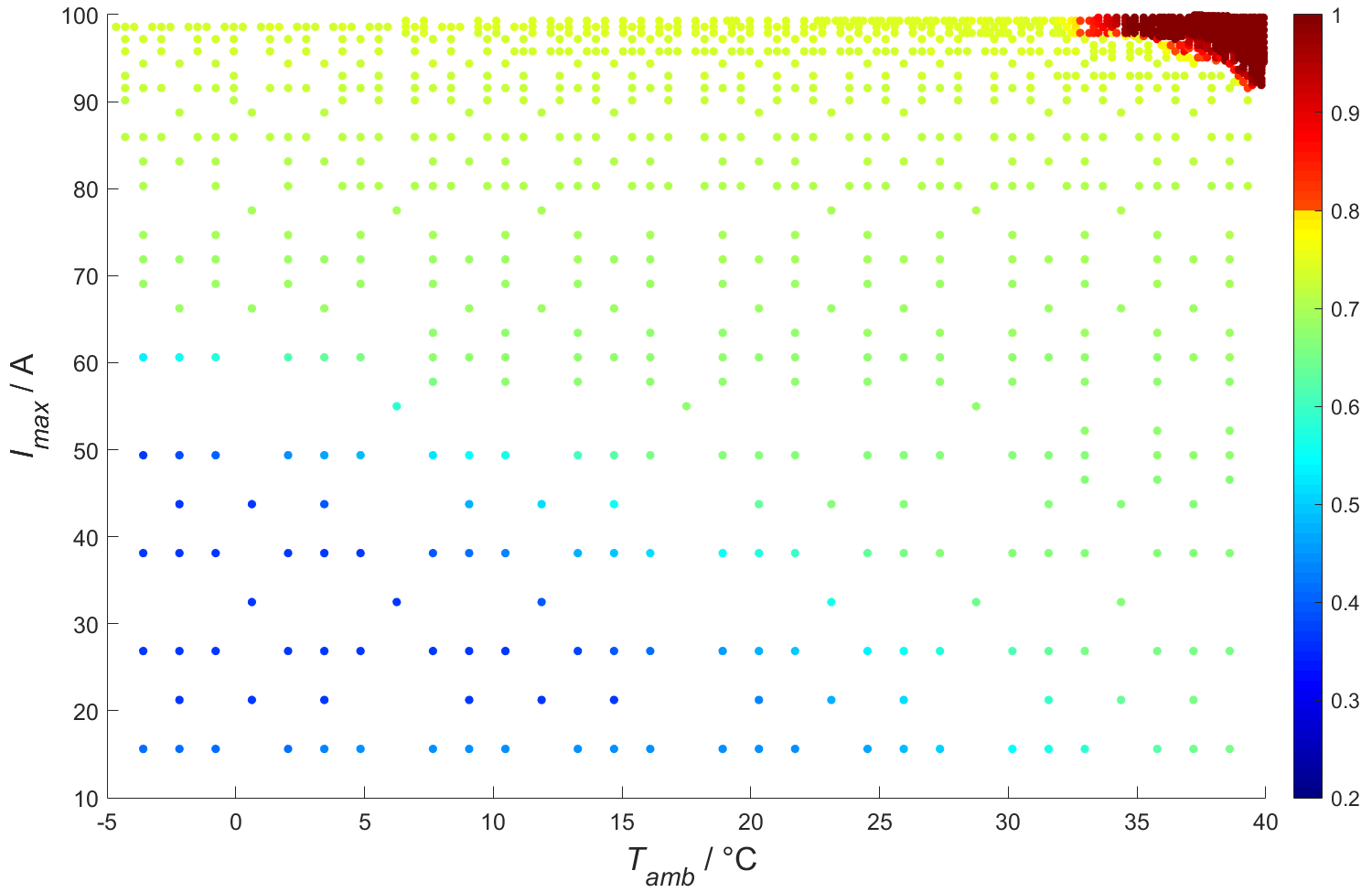}
	\caption{Heat map of the \texttt{SOO} algorithm results with $h_\text{max}(t) = t^\varepsilon$, where $\varepsilon = 0.6$. \textcolor{BrickRed}{Red} marked points indicate a criticality higher than the given threshold $c_\kappa=0.8$.}\label{fig:plot-SOO1}
\end{figure*}

Now we need to find a suitable value for $\varepsilon$. Since $h_\text{max}$ controls the maximum depth of the tree we have to find an $\varepsilon$ that is small enough to actually limit the tree depth forcing the algorithm to explore other tree branches. On the other hand, if $\varepsilon$ is chosen too small, it may happen that there are not enough spaces left in the tree to append new nodes. This is also heavily dependent on the number of children at each node $K$. Low values of $K$ will lead to few spaces in the tree. In our case, we set $K=2$ which means that we cannot set $\varepsilon$ too small so that we choose to vary values between $0.6$ and $0.9$. The results can be found in Table~\ref{tab:SOO-results}. A graphical depiction of the results can be found in Figure~\ref{fig:plot-SOO1}.

\begin{table}[h!]
	\centering
	\begin{tabular}{| l | l |}
		\hline
		\; \textbf{Algorithm} \; & \; \textbf{Critical events} \;\\
		\hline 
		\; \texttt{SOO} ($h_\text{max} = t^{0.6})$ \; & \; $3163$ \;  \\
		\hline 
		\; \texttt{SOO} ($h_\text{max} = t^{0.7})$ \; & \; $3177$ \;  \\
		\hline
		\; \texttt{SOO} ($h_\text{max} = t^{0.8})$ \; & \; $3177$ \;  \\
		\hline
		\; \texttt{SOO} ($h_\text{max} = t^{0.9})$ \; & \; $3177$ \;  \\
		\hline 
	\end{tabular}
	\vspace{0.3cm}
	\caption{Number of rare critical events after $N = 4000$ simulations using \texttt{SOO} with different hyperparameter configurations.}\label{tab:SOO-results}
\end{table}

Overall, we obtain the following results. First, the results do not vary for $\varepsilon>0.7$, meaning that $\varepsilon$ is already so large that $h_\text{max}$ does not limit the tree depth in any meaningful way. The second and more important result is that \texttt{SOO} detects more critical events than \texttt{HOO} and \texttt{POO}, but less than \texttt{DOO}. As the \texttt{SOO} is not given any information about the semi-metric $l$, it has to use more of its simulation budget to explore the space. 

In general, Munos~\cite{munos} suggests to apply the \texttt{DOO} algorithm if the semi-metric $l$ is known, and the \texttt{SOO} algorithm otherwise. In this case study we applied the \texttt{DOO} algorithm with three different parameterizations and found only minor differences. In addition, all three instances detected more events than the \texttt{SOO} algorithm. That is, either all three parameterizations are fairly suited for this use case, or the algorithm is rather robust for this particular use case as the space is structured rather "simply" (the criticality increases with an increasing maximum current and/or ambient temperature).

\section{Discussion}\label{sec:discussion}
The four different algorithms we applied can be classified into two groups. One comprises the \texttt{DOO} and \texttt{SOO} algorithm as both algorithms are developed for the \textit{deterministic} case, i.e., same parameter configurations always lead to same results (which is the case for the model of this use case). For the other two algorithms, \texttt{HOO} and \texttt{POO}, the range of applicability is broader as those can also deal with the \textit{stochastic} case in which the criticality values are disturbed by some type of (bounded) noise.

\begin{figure*}[h!]
	\centering
	\includegraphics[width=\textwidth]{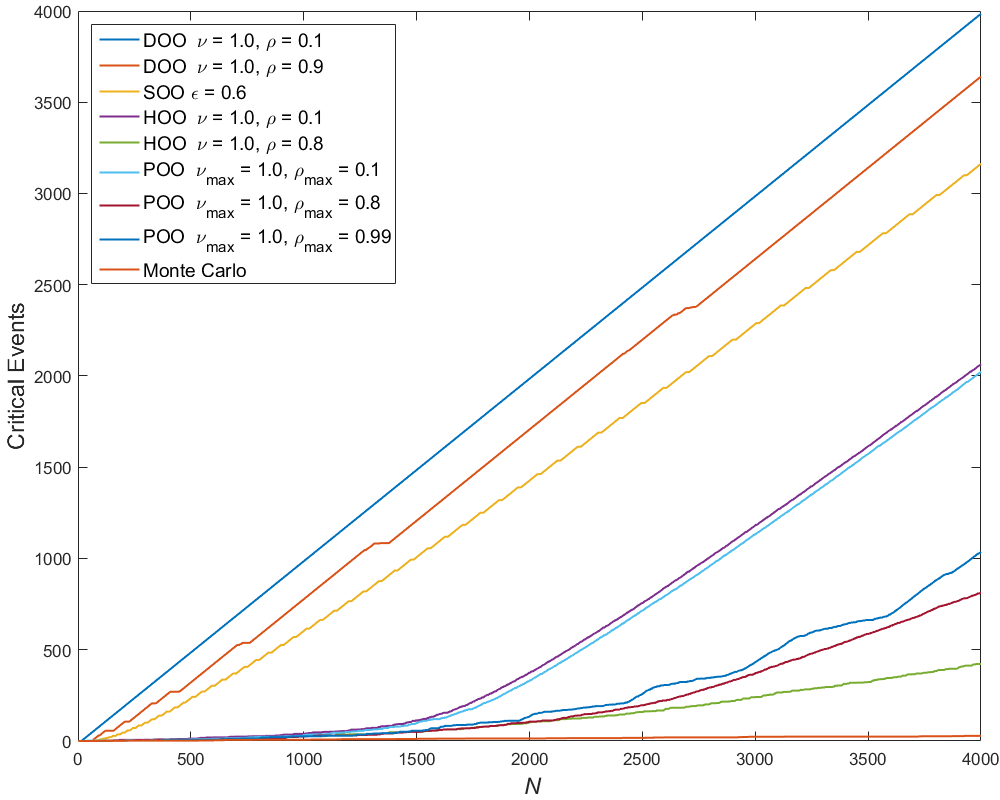}
	\caption{Comparison of the accumulated numbers of detected rare critical events after $N$ simulations for all algorithms applied to the use case.}\label{fig:graphs-all}
\end{figure*}

A graphical comparison of different instances of the algorithms can be found in Figure~\ref{fig:graphs-all}. Overall, the deterministic algorithms (\texttt{DOO} and \texttt{SOO}) performed better than the stochastic algorithms in terms of the number of detected rare events. This is mainly due to the fact that they can fully “trust” in every criticality value measured and do not have to take any possible noise into account. Furthermore, for all tested hyperparameter configurations, all four algorithms performed notably better than Monte Carlo simulation, again in the sense that more critical events have been found. The fact that the algorithms performed better for every tested hyperparameter configuration is a result of particular interest: this indicates the strength of our approach, meaning that even for unfavorable smoothness parameter choices good results can be achieved compared to simple Monte Carlo.  

For each algorithm class (deterministic and stochastic) we also compared an algorithm that requires the knowledge of the smoothness (\texttt{DOO}, \texttt{HOO}) with an algorithm that does not (\texttt{SOO}, \texttt{POO}). Our experiments have shown that for this use case the \texttt{DOO} algorithm outperformed \texttt{SOO} for all hyperparameter combinations tested, despite the fact that the exact degree of smoothness (and semi-metric) is unknown in our case. 

Also the \texttt{HOO} algorithm showed similiar good performances. For a wide range of $\rho$ values, the \texttt{HOO} algorithm performed better than the \texttt{POO} algorithm as shown in Fig. \ref{fig:graphs-HOOandPOOs}. Only for 3 of the 10 instances tested, the \texttt{POO} algorithm performed better. This is also a good indicator for the robustness of the algorithms for this particular use case. Nevertheless, in general we suggest to apply \texttt{POO} or \texttt{SOO} as a first trial on other use cases since this can highly depend on the modeled criticality function, and thus varies from use case to use case.    

Apart from that, in the course of this case study we also found that for very high ambient temperatures and high available currents the requirements stated in Section~\ref{subsec:reqs-and-test-space} are not fulfilled, i.e., that the battery will either overheat or charge unacceptably long. This critical area has been observed with all algorithms that we tested (even with simple Monte Carlo) (see Figs.~\labelcref{fig:plot-MC1,fig:plot-MC2,fig:plot-HOO1,fig:plot-HOO2,fig:plot-DOO1,fig:plot-DOO2,fig:plot-DOO3,fig:plot-SOO1}). The difference between the applied algorithms here lies within the degree of how well this region is covered. For instance, simple Monte Carlo produced only few samples in the critical region (see Figs.~\labelcref{fig:plot-MC1,fig:plot-MC2}). On the other hand, \texttt{DOO} over-sampled the core of the critical region while ignoring the less (but still) critical transition region (see Figs.~\labelcref{fig:plot-DOO1,fig:plot-DOO2,fig:plot-DOO3}). In contrast, the other algorithms (\texttt{HOO}, \texttt{POO}, \texttt{SOO}) delivered a sufficiently precise overview of the region.

One could argue that, due to the fact that the critical area is located at high temperatures and high available current, this critical scenario could have been also found by testing boundary conditions (boundary value analysis), or by testing equivalence classes. In the context of the present use case this might be true but, in general, using the aforementioned methods would (i) not give any information about the actual size of the critical region, and (ii) have a high probability of failure when dealing with more complex parameter spaces, e.g., when the parameter space consists of one or more rather steep peaks centered somewhere in the middle of the space (rare events). In contrast, the algorithms presented in this study have the advantage that their underlying core idea is independent from the location and nature of the critical area as long as the respective assumptions (see Sec.~\ref{sec:methodology}) hold. Note that the difficulty of optimizing the criticality function will still affect their performance (meaning the simulation budget needed to find these rare events).

\section{Conclusion}\label{sec:conclusion}

We tested the Battery Management System provided by AKKA applying simple Monte Carlo simulation and four different approaches from the domain of optimistic optimization within the scope of a simulation-based methodology proposed by OFFIS (see Sec.~\ref{sec:methodology}). The main goal was to showcase the efficacy of that methodology in the sense that (rare) critical events in the system (see Sec.~\ref{sec:testsystem}) are detected with a notably increased frequency compared to simple Monte Carlo simulation. In this context, critical events denote those events for which the given safety requirements (see Sec.~\ref{subsec:reqs-and-test-space}) are violated. 

Overall, regarding the given system under test we found that for very high ambient temperatures and high available currents the requirements are not satisfied. That is, the battery will either overheat or charge unacceptably long for those parameter configurations. When applying simple Monte Carlo (random sampling) we found that this occurs with a frequency of approximately $0.7\%$. In particular, this means that around $99.3\%$ of the entire simulation budget is used for uncritical test runs. With the four algorithms being studied in this report, the probability of sampling in the critical region(s) is notably increased as the results clearly show (see Sec.~\ref{sec:application}), thus meeting the main goal of the case study as stated above. Moreover, we discovered that some of the applied approaches are more suited than others to cover the region of critical parameters with a sufficient precision, i.e. to sketch its shape, at least for the present use case. This might serve as an indicator for the degree of suitability of the different approaches regarding further application contexts, depending on the intended testing objective. 

The presented methodology can make a valuable contribution to safety validation of safety-critical systems. As rare event simulation usually targets estimating the probability of occurrence of rare events effectively, optimistic optimization can be used for this purpose as well. However, this introduces a bias regarding the rare event probability which one has to correct for. Developing a sound mathematical framework for this remains future work. 

\newpage

\bibliographystyle{splncs04}
\bibliography{Case_Study}

\begin{thebibliography}{10}
\providecommand{\url}[1]{\texttt{#1}}
\providecommand{\urlprefix}{URL }
\providecommand{\doi}[1]{https://doi.org/#1}

\bibitem{bartlett18}
Bartlett, P.L., Gabillon, V., Valko, M.: A simple parameter-free and adaptive
  approach to optimization under a minimal local smoothness assumption. In:
  Proceedings of the 30th International Conference on Algorithmic Learning
  Theory (2019)

\bibitem{MasterSim}
{Bauklimatik Dresden}: {MasterSim}. Official website:
  \url{http://bauklimatik-dresden.de/mastersim/}, [Accessed: 22 July 2020]

\bibitem{Bubeck2011}
Bubeck, S., Munos, R., Stoltz, G., Szepesvari, C.: {$\mathcal{X}$}-{A}rmed
  {B}andits. The Journal of Machine Learning Research  \textbf{12},  1655--1695
  (2011)

\bibitem{chang2013}
Chang, W.Y.: The state of charge estimating methods for battery: A review.
  International Scholarly Research Notices  \textbf{2013} (2013)

\bibitem{FMIKit}
{Dassault Systems}: {FMI Kit for Simulink}. Official website:
  \url{https://www.3ds.com/products-services/catia/products/dymola/fmi/},
  [Accessed: 22 July 2020]

\bibitem{Grill2015}
Grill, J.B., Valko, M., Munos, R.: Black-box optimization of noisy functions
  with unknown smoothness. In: Cortes, C., Lawrence, N., Lee, D., Sugiyama, M.,
  Garnett, R. (eds.) Advances in Neural Information Processing Systems 28, pp.
  667--675. Curran Associates, Inc. (2015),
  \url{http://papers.nips.cc/paper/5721-black-box-optimization-of-noisy-functions-with-unknown-smoothness.pdf}

\bibitem{Ismail_2013}
Ismail, N.H.F., Toha, S.F., Azubir, N.A.M., Ishak, N.H.M., Hassan, M.K.,
  Ibrahim, B.S.K.: Simplified heat generation model for lithium ion battery
  used in electric vehicle. {IOP} Conference Series: Materials Science and
  Engineering  \textbf{53},  012014 (dec 2013).
  \doi{10.1088/1757-899x/53/1/012014},
  \url{https://doi.org/10.1088/1757-899x/53/1/012014}

\bibitem{iso26262}
ISO: {Road vehicles -- Functional safety} ({2011})

\bibitem{isopas21448}
ISO: {Road vehicles -- Safety of the intended functionality} ({2019})

\bibitem{Legay2010}
Legay, A., Delahaye, B., Bensalem, S.: Statistical model checking: An overview.
  In: Barringer, H., Falcone, Y., Finkbeiner, B., Havelund, K., Lee, I., Pace,
  G., Ro{\c{s}}u, G., Sokolsky, O., Tillmann, N. (eds.) Runtime Verification.
  pp. 122--135. Springer Berlin Heidelberg, Berlin, Heidelberg (2010)

\bibitem{MATLAB}
{MathWorks}: {MATLAB}. Official website:
  \url{https://de.mathworks.com/products/matlab.html}, [Accessed: 22 July 2020]

\bibitem{Simulink}
{MathWorks}: {Simulink}. Official website:
  \url{https://de.mathworks.com/products/simulink.html}, [Accessed: 22 July
  2020]

\bibitem{Modelica2014}
{Modelica Association}: {Functional Mock-up Interface for Model Exchange and
  Co-Simulation 2.0}.
  \url{https://svn.modelica.org/fmi/branches/public/specifications/v2.0/FMI_for_ModelExchange_and_CoSimulation_v2.0.pdf},
  [Accessed: 22 July 2020] (2014)

\bibitem{Morio2014}
Morio, J., Balesdent, M., Jacquemart, D., Verg{\'e}, C.: A survey of rare event
  simulation methods for static input-output models. Simulation Modelling
  Practice and Theory  \textbf{49},  287--304 (2014)

\bibitem{munos}
Munos, R.: Optimistic {O}ptimization of a {D}eterministic {F}unction without
  the {K}nowledge of its {S}moothness. In: Advances in Neural Information
  Processing Systems. pp. 783--791 (2011)

\bibitem{murnane2017}
Murnane, M., Ghazel, A.: {A Closer Look at State of Charge (SOC) and State of
  Health (SOH) Estimation Techniques for Batteries}. Analog devices
  \textbf{2},  426--436 (2017)

\bibitem{nicolai2018co}
Nicolai, A.: {Co-Simulations-Masteralgorithmen -- Analyse und Details der
  Implementierung am Beispiel des Masterprogramms MASTERSIM}.
  \url{http://bauklimatik-dresden.de/mastersim/2nd/doc/Nicolai_Qucosa_2018_MasterSim_Algorithmus.pdf},
  [Accessed: 22 July 2020] (2018)

\bibitem{okelly2018}
O\textquotesingle~Kelly, M., Sinha, A., Namkoong, H., Tedrake, R., Duchi, J.C.:
  Scalable end-to-end autonomous vehicle testing via rare-event simulation. In:
  Bengio, S., Wallach, H., Larochelle, H., Grauman, K., Cesa-Bianchi, N.,
  Garnett, R. (eds.) Advances in Neural Information Processing Systems.
  vol.~31. Curran Associates, Inc. (2018),
  \url{https://proceedings.neurips.cc/paper/2018/file/653c579e3f9ba5c03f2f2f8cf4512b39-Paper.pdf}

\bibitem{piller2001}
Piller, S., Perrin, M., Jossen, A.: Methods for state-of-charge determination
  and their applications. Journal of Power Sources  \textbf{96}(1),  113--120
  (2001)

\bibitem{Puch2019diss}
Puch, S.: Statistisches Model Checking mittels gef{\"{u}}hrter Simulation im
  Kontext modellbasierter Entwicklung sicherheitskritischer
  Fahrerassistenzsysteme. Ph.D. thesis, Carl von Ossietzky Universit{\"{a}}t
  Oldenburg (2019)

\bibitem{zhao17}
Zhao, D., Huang, X., Peng, H., Lam, H., LeBlanc, D.J.: Accelerated {E}valuation
  of {A}utomated {V}ehicles in {C}ar-{F}ollowing {M}aneuvers. IEEE Transactions
  on Intelligent Transportation Systems  \textbf{19}(3),  733--744 (2017)

\bibitem{zhao16}
Zhao, D., Lam, H., Peng, H., Bao, S., LeBlanc, D.J., Nobukawa, K., Pan, C.S.:
  Accelerated {E}valuation of {A}utomated {V}ehicles {S}afety in
  {L}ane-{C}hange {S}cenarios based on {I}mportance {S}ampling {T}echniques.
  IEEE Transactions on Intelligent Transportation Systems  \textbf{18}(3),
  595--607 (2016)

\end{thebibliography}

\end{document}